\documentclass{article}

\usepackage{libertine}
\usepackage{amsthm, amsmath, mathtools}
\usepackage{amssymb}
\usepackage{xcolor}
\usepackage{algorithm}
\usepackage{algorithmicx}
\usepackage[noend]{algpseudocode}
\usepackage{comment}
\usepackage{enumerate}
\usepackage{multicol}
\usepackage[normalem]{ulem}
\usepackage{xspace}
\usepackage[format=plain,
            labelfont=it,
            textfont=it]{caption}

\usepackage{geometry}
\usepackage{color}
\usepackage{phonetic}
\usepackage{tipa}
\usepackage{bbm}
\usepackage{pifont}
\usepackage{wasysym}
\usepackage{enumitem}
\usepackage{marvosym}
\usepackage{sidecap}
\usepackage[labelfont=bf, textfont=up]{caption}

\usepackage[normalem]{ulem}

\usepackage[capitalise]{cleveref}

\usepackage{lineno}
\usepackage{balance}

\usepackage{latexsym}

\usepackage{txfonts}
\usepackage{authblk}

\newtheorem{lemma}{Lemma}
\newtheorem{definition}[lemma]{Definition}

\newtheorem{theorem}{Theorem}

\newcommand{\Pcom}[1]{{\color{orange}#1}}

\newcommand{\Pst}[1]{\sout{#1}}

\newcommand{\JMcom}[1]{{\color{green}#1}}

\newcommand{\JCom}[1]{{\color{red}#1}}

\newcommand{\Jst}[1]{\sout{#1}}

\renewcommand{\Pcom}[1]{}

\renewcommand{\Pst}[1]{}

\renewcommand{\JMcom}[1]{}

\renewcommand{\JCom}[1]{}

\renewcommand{\Jst}[1]{}

\crefname{observation}{Observation}{Observations}
\crefname{corollary}{Corollary}{Corollaries}

\title{Bag Semantics Conjunctive Query Containment.\\ Four Small Steps Towards Undecidability\footnote{This research  was  supported by grant
  2022/45/B/ST6/00457 from the Polish National Science Centre (NCN).
}}

\author{\vspace{4mm}Jerzy Marcinkowski ~~~~~~~~~~~~~~~ Mateusz Orda}
\affil{\vspace{4mm}\normalsize Institute of Computer Science, University of Wrocław; Wrocław, Poland.}
\date{}

\begin{document}

\maketitle

\begin{abstract}
Query Containment Problem (QCP) is one of the most
fundamental  decision problems in database query processing and optimization.

Complexity of QCP for conjunctive queries ($QCP_{\text{\tiny CQ}}$) has been fully understood since 1970s.
But, as Chaudhuri and Vardi noticed in their classical paper  \cite{CV93},
this understanding is based on the assumption that  query answers are
sets of tuples, and it does not transfer to the situation when multi-set (bag) semantics
is considered.

Now, 30 years after \cite{CV93} was written, decidability of  $QCP_{\text{\tiny CQ}}$ for bag semantics remains an open question,
 one of the most intriguing open questions in database theory.

In this paper we show a series of undecidability results for some generalizations of bag-semantics $QCP_{\text{\tiny CQ}}$.
We show, for example, that the problem whether, for given two boolean conjunctive queries $\phi_s$ and $\phi_b$,
and a linear function $\mathbb F$, the inequality ${\mathbb F}(\phi_s(D))\leq \phi_b(D)$ holds for each database instance $D$,
is undecidable\footnote{Notice that if $\phi$ is a boolean conjunctive query then, under the multiset semantics, $\phi(D)$ is a natural number.}.
\end{abstract}

\newcommand{\Arena}{{\text \footnotesize{ Arena}}}
\newcommand{\ddd}{\mathbbmss{d}}
\newcommand{\nnn}{\mathbbmss{n}}
\newcommand{\mmm}{\mathbbmss{m}}
\newcommand{\ccc}{\mathbbmss{c}}
\newcommand{\qqq}{\mathbbmss{q}}

\renewcommand{\lll}{\mathbbmss{l}}

\newcommand{\kkkk}{{\text \texthtk}}

\newcommand{\cccc}{  {\text \texthtc} }
\newcommand{\llll}{ \textbeltl}
\newcommand{\jjjj}{ {\text \textctj }}
\newcommand{\mmmm}{\text{\textltailm}}
\newcommand{\nnnn}{\text{\textltailn}}


\newcommand{\cqneg}{{CQ_{\neq}}}

\newcommand{\wedgebar}{\hspace{0.5mm}\bar{\wedge}\hspace{0.5mm}}

  \renewcommand{\clubsuit}{ $\ding{95}$ }
  
 \renewcommand{\spadesuit}{ $\ding{96}$ }

  \renewcommand{\;}{ \thickspace }

\newcommand{\xxi}{\mathbbmss x }

\section{The context. And our contribution.}

\subsection{The context}\label{intro-1}

Query Containment Problem (QCP) is one of the most  fundamental  decision problems in database query processing and optimization. 
It is formulated as follows:\medskip

\noindent
{\em The instance of QCP} are two database queries, $\Psi_s$ and $\Psi_b$.

\noindent
{\em The question} is whether 
$\Psi_s(D)\subseteq \Psi_b(D)$ holds for each database $D$.\medskip

In the above, by $\Psi(D)$  we denote\footnote{See Section \ref{sec:prelims} for more explanation regarding the notations we are using.}
the result of applying query $\Psi$ to the database $D$.
If the reader wondered what the  subscripts $s$ and $b$ are supposed to mean:
$s$ stands for ,,small'' and $b$ stands for ,,big'' (and we use this naming convention through this paper, sometimes also using terms $s$-query and $b$-query for  $\Psi_s$ and   $\Psi_b$ ). QCP asks if the answer to the ,,small'' query is always contained in the answer to the ,,big'' one.

As usual in such situations, the problem comes in many variants, depending on two parameters: on the class of queries we allow and on the
precise semantics of  $\Psi(D)$ (and -- in consequence -- of the precise semantics of the symbol $\subseteq$). 
The classes of queries which have been considered in this context are
$CQ $ (conjunctive queries), or $UCQ$ (unions of conjunctive queries) or $\cqneg$ (conjunctive queries 
with inequalities), or some subsets of $CQ$. The possible semantics of $\Psi(D)$ are two: either we can see $\Psi(D)$ as a relation, 
that is a {\bf set} of tuples, or as a multirelation, 
that is a {\bf multiset} also known as  a {\bf bag} of tuples\footnote{We should probably remark here that,
while $\Psi(D)$ may be a multiset,
$D$ is always a
relational structure in this paper. }. In the first case, the $\subseteq$ in the above statement
of QCP is understood to be the
set inclusion, in the second case it is the multiset inclusion.
We use natural notations to call the variants, for example $QCP^{\text{\footnotesize bag}}_{\text{\tiny CQ,} \tiny \neq}$ is
QCP for conjunctive queries with inequality, under bag semantics and $QCP^{\text{\footnotesize set}}_{\text{\tiny UCQ}}$ is $QCP$ for unions of CQs  under set semantics.

 $QCP^{\text{\footnotesize set}}$ has long been well understood. Already in 1977
 Chandra and Merlin \cite{CM77} realized that $QCP^{\text{\footnotesize set}}_{\text{\tiny CQ}}$ is NP-complete.
 Concerning more general classes of  queries,
 it was shown in  \cite{SY80} that $QCP^{\text{\footnotesize set}}_{\text{\tiny UCQ}}$ is $\Pi^{\text{\tiny P}}_2$-complete
. Then another more general class,
conjunctive queries with
comparison predicates $\neq$ and $\leq$, was studied in \cite{K88}, where it was proven that
 $QCP^{\text{\footnotesize set}}_{\text{\tiny CQ},\neq,\leq}$   is also in  $\Pi^{\text{\tiny P}}_2$,  but no lower bound was established.
This gap was finally filled by  \cite{M97}, which proves that $QCP^{\text{\footnotesize set}}_{\text{\tiny CQ},\neq,\leq}$   is
$\Pi^{\text{\tiny P}}_2$-complete.

But a  case can be made that
in real database systems, 
where duplicate tuples are not eliminated,
queries are usually evaluated under bag semantics, not set semantics.

Unfortunately, as it was realized in the early 1990s,
no tools or techniques developed for the analysis of  $QCP^{\text{\footnotesize set}}$ survive in the context of  $QCP^{\text{\footnotesize bag}}$.
In the seminal paper
 \cite{CV93} the authors observe that
the proof of the Chandra-Merlin NP upper bound  for
$QCP^{\text{\footnotesize set}}_{\text{\tiny CQ}}$
 does not
survive in the bag-semantics world, and claim a $\Pi^P_2$ lower bound for  $QCP^{\text{\footnotesize bag}}_{\text{\tiny CQ}}$,
deferring the proof however to the full version of the paper.

The same observation was made also in an earlier\footnote{It seems that the authors of \cite{CV93} were not aware of \cite{CW91}.} paper \cite{CW91}, less well known than \cite{CV93}. Let us quote \cite{CW91} here: {\em The classical theorem by Chandra and Merlin does not hold, because
it treats relations as sets and not multisets. (...). In general, there is almost no
theory on the properties of queries and programs that retain duplicates. The
development of such a theory is part of our future plans.}

But such theory was never really developed, the full version of \cite{CV93} never appeared, and
neither the authors of  \cite{CV93}, nor anyone later on, proved any upper bound\footnote{We of course know that the problem is in co-r.e.} for the complexity of $QCP^{\text{\footnotesize bag}}_{\text{\tiny CQ}}$.
So not only nothing is known about the complexity of this fundamental problem but even its decidability has now been an open problem for 30 years. And this is not because people did not try.

When  a difficult decision problem is attacked, the action usually takes place in two theaters of operations:
on the positive side, where more and more general subcases of the  problem are being proven to be decidable, 
and on the negative side, where undecidability results are shown for some generalizations of the problem. 

On the {\bf positive side}, numerous results were produced, which seem  to naturally fall into two main lines of attack.

 One of these lines includes 
  decidability of $QCP^{\text{\footnotesize bag}}$ for
projection-free conjunctive queries  \cite{ADG10}.
 In a related paper   \cite{KRS12} the authors generalize QCP and then give a partial positive answer
 for the problem of query containment of (unions of)
conjunctive queries over so called annotated databases.
This line of research was continued in \cite{cohen09} and \cite{chirkova12}.Then,
more recently, the decidability result from \cite{ADG10} was extended to the case where
$\Psi_s$ is a projection-free CQ and $\Psi_b$ is an arbitrary CQ \cite{KM19}.
The proof is via a reduction to a known decidable class of Diophantine inequalities.

The second line of attack  originated from the work of
Kopparty and Rossman \cite{KoR11}.
 They observe that $QCP^{\text{\footnotesize bag}}_{\text{\tiny CQ}}$  is a purely combinatorial (or graph theoretic)  phenomenon
related to the notion of homomorphism domination exponent. In consequence, they postulate that the
existing combinatorial technology could be used to approach the problem
(and indeed they prove decidability for a case  when $\Psi_s$ is  series-parallel and $\Psi_b$ is a chordal graph). Their mathematically very attractive
toolbox
features    the information-theoretic notion of entropy.
Unfortunately, the paper \cite{AKNS20}  exhibits the limitations of this attitude, showing that
decidability of $QCP^{\text{\footnotesize bag}}_{\text{\tiny CQ}}$, even if restricted to the case where  $\Psi_b$ is an acyclic CQ,
is already equivalent to an
 long standing open problem in information theory,  decidability of the problem Max-IIP.\

On the {\bf negative side}, which is more interesting from the point of view of our paper, the results are so far very few. All we know is that
the two most natural extensions of $QCP^{\text{\footnotesize bag}}_{\text{\tiny CQ}}$ are undecidable. First  \cite{IR95}
proved that  $QCP^{\text{\footnotesize bag}}_{\text{\tiny UCQ}}$ is undecidable. The proof is quite easy -- it is a straightforward encoding of Hilbert's 10th problem. Then, in 2006, \cite{JKV06} have shown  that $QCP^{\text{\footnotesize bag}}_{\text{\tiny CQ},\neq}$  is
also undecidable. The argument here is much more complicated than the one in \cite{IR95}  and, while ``real'' conjunctive queries are mentioned in the title of \cite{JKV06},
the queries needed for the proof of this negative result require no less than $59^{10}$ inequalities.

No progress has been made in this theater of operations since that time.

  \subsection{Our contribution} \label{twierdzenia}
  
In this paper we present a series of negative results for some generalizations of  $QCP^{\text{\footnotesize bag}}_{\text{\tiny CQ}}$. We
also notice that (some of) our results are (in some sense) ultimate: no stronger undecidability result is possible unless the problem $QCP^{\text{\footnotesize bag}}_{\text{\tiny CQ}}$
 itself is undecidable. 

 All our results hold for {\bf boolean conjunctive queries} (with, or without, inequality).
 For a boolean query, the result of its application to a database is a natural number (see  Section \ref{sec:prelims}) and, in consequence,
 the $\subseteq$ symbol from the QCP statement at the beginning of Section \ref{intro-1} turns into $\leq$.

Call a database $D$ {\bf non-trivial} if it contains two {\bf different} constants, $\mars$ and $\venus$.
  
  Our first result is:

  \begin{theorem}\label{th-z-c}
  {\em The problem:}\\
  Given are boolean conjunctive queries (without inequality)  $ \phi_s $ and $\phi_b $, and a natural number $\ccc$.
  Does $\ccc\phi_s(D) \leq \phi_b(D)$ hold for each  non-trivial database $D$?\\
{\em is undecidable.}
\end{theorem}

Notice that Theorem \ref{th-z-c} would make no sense
without the condition that $D$ must be non-trivial, at least not for any $\ccc>1$.
This is because if $D$ is the ``well of positivity'' --  a structure with a single
vertex\footnote{For a similar  reason, the homomorphism domination exponent in \cite{KoR11} is  only defined for structures
which allow for at least two different homomorphisms, which also rules out single-vertex databases.}
such that all atomic formulas are true in $D$ for this vertex then whatever queries $ \phi_s $ and $\phi_b $ (without inequality) we take,
we get
 $\phi_b(D)=\phi_b(D)=1$, and thus $\ccc\phi_s(D)>\phi_b(D)$.

 As the following theorem says we can however replace the non-triviality condition with an additive constant:
 it is undecidable, for two conjunctive queries $ \phi_s $ and $\phi_b$, without  equality, and a linear function $\mathbb F$, whether
 ${\mathbb F}(\phi_s(D))$ is bounded by $\phi_b(D)$. In other words:
 
 \begin{theorem}\label{th-z-c-c}
  {\em The problem:}\\
  Given are boolean conjunctive queries (without inequality)  $ \varphi_s $ and $\varphi_b $, and  natural numbers  $\ccc$ and  $\ccc'$.
  Does $\ccc\varphi_s(D) \leq \varphi_b(D)+\ccc'$ hold for each   $D$?\\
{\em is undecidable.}
\end{theorem}

 Our next result does not involve any multiplicative constants:

  \begin{theorem}\label{th-po-jednej-nier}
  {\em The problem:}\\
 Given are boolean conjunctive queries (without inequality)  $ \psi_s $ and $\psi_b $ (with at most one inequality).
  Does  $\psi_s(D)\leq \psi_b(D)$  hold for each  non-trivial database $D$ ?\\
{\em is undecidable.}
\end{theorem}

Non-triviality can of course be enforced by replacing $\psi_s$ with $\mars\neq \venus \wedge \psi_s$. So Theorem \ref{th-po-jednej-nier} is an improvement upon
the main result from \cite{JKV06}:
we get undecidability of $QCP^{\text{\footnotesize bag}}_{\text{\tiny CQ},\neq}$ already for queries with one inequality each, instead of $59^{10}$.

Notice that again the non-triviality assumption cannot be easily dropped here:  by the ``well of positivity'' argument the query
$\rho_b\wedge (x\neq x')$ never contains $ \rho_s $ (for queries $ \rho_s $ and $\rho_b $ being CQs without equality).
The next theorem  shows, that this ``well of positivity'' argument is actually the only reason why
we need non-triviality:

  \begin{theorem}\label{plusik-jeden}
  {\em The problem:}\\
 Given boolean conjunctive queries $ \rho_s $ (without inequality)   and $\rho_b $  (with at most one inequality).
  Does  $\rho_s(D)\leq max\{1, \rho_b(D)\}$  hold for each  database $D$?\\
 {\em is undecidable.}
\end{theorem}

Theorem \ref{plusik-jeden} says (with  one caveat) that the inequality in the $s$-query is not really essential for the negative result
from \cite{JKV06}.
Finally, we tried to address the question, whether we could  get rid of the inequality in the $b$-query rather than the one on the
$s$-query and still prove undecidability.
The answer is given in Theorem \ref{dodatkowe}: maybe we can, but
 this can only happen if $QCP_{\text{\tiny CQ}}^{\text{\footnotesize bag}}$ itself is undecidable:

\begin{theorem}\label{dodatkowe}
 {\em The problem:}\\
Given two conjunctive queries  $ \psi_s $ (with any number of inequalities) and $\psi_b $ (without inequalities),
  Does  $\psi_s(D)\leq\psi_b(D)$ hold for each database $D$?\\
{\em is decidable if and only if $QCP_{\text{\tiny CQ}}^{\text{\footnotesize bag}}$ is decidable.}
\end{theorem}

  What concerns {\bf our techniques:} all the undecidability results in bag-semantics database theory we are aware of 
  (\cite{IR95}, \cite{JKV06}, \cite{KMO22}) use Hilbert's 10th Problem as the source of undecidability 
  (see Section \ref{source-of-u}). In all these papers 
  the database provides a valuation of the numerical variables, and the universal quantification from the Hilbert's Problem is 
  simulated by the universal quantification over databases. The question is
  how to encode the evaluation of a given polynomial using the syntax under analysis.

  This  can be easily done if we deal with UCQs, like in \cite{IR95}: a monomial translates in a very natural way  into a CQ, and a sum of monomials into a disjunction of CQs.
  
  In \cite{JKV06} a trick was found to encode an entire polynomial as one CQ. But this only works well for some ``good'' databases. So this
 trick has been married in \cite{JKV06} to an elaborate anti-cheating mechanism, which guarantees that if $D$ is not ``good'' then 
 (using the language of Theorem \ref{th-po-jednej-nier} above) $\psi_b(D)$ is easily big enough to be greater than $\psi_s(D)$. 
  And it is this anti-cheating mechanism in  \cite{JKV06} that
  requires an astronomical number of inequalities. 
  
  The main idea behind our proof of Theorems \ref{th-z-c} and \ref{th-z-c-c}  is a new polynomial-encoding trick, which superficially looks quite similar to the one from
  \cite{JKV06}, and also only works for ``good databases'' (which we call {\em correct} in Section \ref{sec:glowne-dowod-1}), but is different enough
  not to require any inequalities in the anti-cheating part (in particular, 
 nothing similar to Lemma \ref{trik-ktorego-oni-nie-widzieli}, which is very important in our proof, seems to be compatible
  with the trick from  \cite{JKV06}). This  comes with a cost however, which is the multiplicative constant $\ccc $.

 Theorems \ref{th-po-jednej-nier} and \ref{plusik-jeden} follow from Theorem \ref{th-z-c}.
 We use a combinatorial argument to show how to use
 the single inequality in $\psi_b$ to
 simulate multiplication by $\ccc$. This part is not related to any previous work we are aware of. 

{\bf Organization of the paper.} In Section \ref{sec:mnozenie} we show how Theorem \ref{th-po-jednej-nier} follows from Theorem \ref{th-z-c}.
Theorem \ref{th-z-c} itself is proved in Section \ref{sec:glowne-dowod-1} (due to space limitation proofs of two lemmas are presented in  Appendix A and Appendix B). Sections \ref{sec:mnozenie} and  \ref{sec:glowne-dowod-1} can be read in any order.

Proofs of Theorems \ref{plusik-jeden} and \ref{th-z-c-c}   are  deferred to  the full paper.
They follow the same paths as the proofs of Theorems \ref{th-z-c} and \ref{th-po-jednej-nier}, but there are some new obstacles there to overcome.
This is because the anti-cheating mechanism in the proofs of Theorems \ref{th-z-c} and \ref{th-po-jednej-nier}
strongly relies on the input database to be non-trivial. Which means that another level of anti-cheating response must be
present, in the proofs of Theorems \ref{plusik-jeden} and \ref{th-z-c-c}, to
make sure that in trivial databases the answer to the $b$-query is always at least big as  the answer to the $s$-query.

The (easy) proof of Theorem \ref{dodatkowe} can be found in Section \ref{dyskusja}.

We begin the technical part with Section \ref{sec:prelims} where our notations and other basic concepts are explained.


\section{Preliminaries and notations}\label{sec:prelims}

\subsection{Some standard notions and notations}\label{notions-and-notations}
We use the terms {\bf database} and {\bf structure} interchangeably, to denote 
a finite relational structure over some relational schema (signature). Apart from relations we allow
 for constants (see Section \ref{o-stalych}) in the signature: $a$ and $b$ (possibly with  subscripts) are  used to denote such constants
 as well as $\mars$ and $\venus$. We  use the letter $D$ (possibly with  subscripts) to denote
 structures. If $D$ is a structure then by $V_D$ we mean the set of vertices of $D$ (or the active domain of $D$ 
 if you prefer a more database-theoretical terminology). If $a$ is a constant and $D$ is a structure then we also use
 $a$ to denote the interpretation of $a$ in $D$ (instead of the more formally correct $a_D$).
 
 When we say {\bf \em ``query''} we {\bf always}\footnote{With the only exception for Section \ref{intro-1}, where
more general classes of queries are discussed, and where we do not assume that queries are boolean. To distinguish,
we use the upper case Greek letters to denote such more general queries.}  mean a conjunctive query  (CQ).
{\bf All} queries in this paper are {\bf boolean}. We never explicitly
write
the existential quantifiers if front of queries, but
whenever we write a query it is assumed that all the variables are
existentially quantified. To denote queries we always use lower case Greek letters. 
If $\psi$ is a query then $Var(\psi)$ is the set consisting of all the variables which appear in $\psi$ and $V_\psi$
 is the set consisting of all the variables {\bf and}
all the constants which appear in $\psi$. This is consistent with $V_D$ being the active domain of a database, since
we  tacitly identify queries with their {\bf canonical structures}: the active domain of the canonical structure of
query $\psi$ is $V_\psi$ and the atoms of the canonical structure are the atoms of $\psi$.

Queries may contain {\bf inequalities}: by inequality we mean an atomic formula of
the form $x\neq x'$, where $x$ and $x'$ are variables or constants. We think that 
$\neq $ is a binary relation symbol which, for each structure $D$, is interpreted in $D$ as 
the relation $(V_D\times V_D) \setminus \{[s,s]:s\in V_D\}$.

For two structures $D$ and $D'$ by $Hom(D,D')$ we denote the set of all {\bf homomorphisms} from $D$ to $D'$. Note that 
if $a$ is a constant of the language and $h\in Hom(D,D')$ is a homomorphism then $h(a)=a$. We always use $h$ and $g$ to 
denote homomorphisms.

If $\psi$ is a query and $D$ is a structure then $\psi(D)$ denotes the result of the application of $\psi$ to
$D$. Since $\psi$ is boolean, one would  intuitively think that the result can only be {\sc YES} or {\sc NO}, but since we 
consider the multiset (bag) semantics in this paper, this {\sc YES} can be repeated any positive natural number of times,
depending on the number of ways $\psi$ can be satisfied in $D$, and the above intuition is formalized as:
~
$$\psi(D)\stackrel{\text{\tiny df}}{=} |Hom(\psi, D)|$$

Notice that $\psi(D)$ is always a natural number, so in particular we can multiply it by a constant, as we do in Theorem \ref{th-z-c}.
We use the notation $D\models \psi$ if the set $Hom(\psi, D)$ is nonempty.

Notice also that the above definition of $\psi(D)$  makes perfect sense for queries with inequalities.  This is
since if $x\neq x'$ is an inequality in $\psi$ and if $h\in Hom(\psi,D)$ is a homomorphism then:
~
$$ [h(x), h(x')]\in V_D\times V_D \setminus \{[s,s]:s\in V_D\} $$

which means that $x$ and $x'$ are indeed mapped by $h$ to different elements of $V_D$.

\subsection{Disjoint conjunction and query exponentiation}

In our proofs
we often construct conjunctive queries as conjunctions of smaller conjunctive queries. When,  for two conjunctive queries $\rho$ and $\rho'$, we write
$\rho \wedge \rho'$, we think that
the conjunction of the quantifier-free parts of $\rho$ and $\rho'$ is taken first, and then the result is existentially quantified.

If we want to treat the variables in $\rho$ and $\rho'$ as local (which amounts to assuming that
the existential quantification came first, and conjunction later)
we write $\rho \wedgebar \rho'$ instead of  $\rho \wedge \rho'$.

Obviously:

\begin{lemma}
For each $D$, $\rho$ and $\rho'$ it is: $(\rho \wedgebar \rho')(D)=\rho(D)\rho'(D)$.
\end{lemma}

It is equally obvious that the lemma would not be true if, in its statement,  we replaced $\wedgebar$ with $\wedge$.

The symbol $\bar \bigwedge$ relates to $\wedgebar$ as $\bigwedge$ relates to $\wedge$:

\begin{definition}
For $k\in \mathbb N$
by 
$ \theta\uparrow k $ 
we denote
the query $\bar \bigwedge_{1\leq i \leq k}\; \theta$.
\end{definition}

Clearly, for any database $D$, query $\theta $, and number $k\in \mathbb N$ we have:
$(\theta\uparrow k)(D)= (\theta(D))^k$.

\subsection{Short remark about the role of constants}\label{o-stalych}

As we said in Section \ref{notions-and-notations}, all the queries we consider in this paper are boolean, but we allow
for constants in the language. Can we somehow get rid of the constants, possibly allowing non-boolean queries instead?

Imagine $\phi_s$ and $\phi_b$, boolean queries, with some tuple  $\bf a$ of constants. And let $\phi'_s$ and $\phi'_b$ be
syntactically the same queries, but now $\bf a$ is understood to be a tuple of  variables. Free variables. Then the
observation is (and excuse us if it is too obvious) that:
$\phi_b$ contains $\phi_s$ if and only if $\phi'_b$ contains $\phi'_s$.

This is true for any semantics (set or multiset). And this is also true if $\bf a$ are not {\em all } the constants that occur in
 $\phi_s$ and $\phi_b$ but only some of them.

 With the above observation in mind let us see what happens with our results if we ban constants. Since the non-triviality condition is
 important for us, and to express this condition we need constants, there are two possible versions of such ban: soft one, where
 all constants {\em except for} $\mars$ and $\venus$ are disallowed, and hard, where $\mars$ and $\venus$ are disallowed too.

 In the soft version, Theorems \ref{th-z-c} and \ref{th-po-jednej-nier} survive almost intact (after one defines, in the natural way, what $\ccc\cdot M$ means, for a natural number
 $\ccc$ and a multiset $M$). In the hard version, Theorem \ref{th-po-jednej-nier} survives, but with the additional inequality $\mars\neq \venus$ in the $s$-query.


\section{How to multiply. From Theorem \ref{th-z-c} to Theorem \ref{th-po-jednej-nier}. }\label{sec:mnozenie}

In this section we show how Theorem \ref{th-po-jednej-nier} follows from Theorem \ref{th-z-c} (as stated in Section \ref{twierdzenia}).
To this end, suppose we have given $ \phi_s $ and $\phi_b $, like in Theorem \ref{th-z-c}, both CQs without inequality,
and a natural number $\ccc$.

\begin{definition}\label{def:mnozenie}
For a rational number $\qqq>0$ we say that CQs (with or without inequality)
$\varrho_s$ and $\varrho_b$ {\em multiply by $\qqq$} if:\medskip

\noindent
($=$)\hspace{3mm} there exists a non-trivial database $D$ such that $\varrho_s(D)=\qqq\varrho_b(D)\neq 0$;

\noindent
($\leq$) \hspace{2mm} for each non-trivial database $D$ it holds that  $\varrho_s(D)\leq \qqq\varrho_b(D)$.
\end{definition}

In order to prove Theorem \ref{th-po-jednej-nier}
it will now be enough
to construct conjunctive queries  $ \alpha_s $ (without inequalities) and $\alpha_b $ (with at most one inequality)
which multiply by $\ccc$.

Indeed, suppose we have such $\alpha_s$ and $\alpha_b$, whose schema is disjoint from the schema of $\phi_s$ and $\phi_b$.
 Define $\psi_s$ as $ \alpha_s \wedgebar \phi_s$ and $\psi_b$ as   $\alpha_b \wedgebar \phi_b$.
Then it is not hard to notice that the following two conditions are equivalent:

\noindent
 (i)  there exist a non-trivial  database $D$, such that $\ccc\phi_s(D)>\phi_b(D)$;\\
  (ii) there exist a non-trivial database $D$, such that $\psi_s(D)>\psi_b(D)$.\medskip

  Indeed, for the proof that (i) $\Rightarrow$ (ii), let $D_1$ be such a database, over the schema of $\phi_s$ and $\phi_b$, that
   $\ccc\phi_s(D_1)>\phi_b(D_1)$ and let $D_2$ be such a database, over the schema of $\alpha_s$ and $\alpha_b$, that
   $\alpha_s(D_2)=\ccc\alpha_b(D_2)$. Let also $D=D_1\cup D_2$. Then:
~
   $$\ccc \psi_s(D)= (\ccc \phi_s(D_1))\alpha_s(D_2) > \phi_b(D_1)(\ccc \alpha_b(D_2))= \ccc \psi_b(D)$$
~
 In order to see the  $\neg$(i) $\Rightarrow\neg$(ii) implication, suppose that for each non-trivial $D$ it holds that
 $\ccc\phi_s(D)\leq \phi_b(D)$. We also know that for each non-trivial $D$ the inequality $\alpha_s(D)\leq \ccc\alpha_b(D)$ holds.
 This implies that for each $D$:
~
$$(\ccc\phi_s(D))\alpha_s(D) \leq \phi_b(D)(\ccc\alpha_b(D))$$
~
 which means that $\psi_s(D)\leq\psi_b(D)$.

In the next two subsections we will construct $ \alpha_s $ and $\alpha_b $, as specified above. But before we go there,
notice that:

\begin{lemma}\label{widacprzeciez}
Suppose $\varrho_s$ and $\varrho_b$ multiply by some $\qqq$ and $\varrho'_s$ and $\varrho'_b$ multiply by some $\qqq'$.
Assume also that the schema of  $\varrho_s$ and $\varrho_b$ is disjoint from the schema of $\varrho'_s$ and $\varrho'_b$.
Then $\varrho_s\wedgebar \varrho'_s$ and $\varrho_b \wedgebar \varrho'_b$ multiply by $\qqq\qqq'$.
\end{lemma}

\subsection{The workhorse: queries  $ \beta_s $ and $\beta_b $}\label{workhorse}

As we said above, the plan is now to construct conjunctive queries, $\alpha_s$ (with no inequalities) and $\alpha_b$
(with a single inequality) which multiply by $\ccc$. But is it possible at all\footnote{Let us remark here, that a pair of conjunctive queries
without inequality cannot multiply by a number greater that 1.} for such queries to multiply by a number
greater than 1? And (if so), is it possible for such queries to multiply by an arbitrarily  huge number? This is what this subsection is about.

Let $R$ be a new relational symbol of some arity $\nnnn \geq 3$. Define query
$CYCLIQ(x_1, x_2,\ldots x_{\nnnn})$ as:
$$R(x_1, x_2,\ldots x_{\nnnn})\wedge R(x_2,\ldots x_{\nnnn}, x_1)\wedge \ldots \wedge R(x_{\nnnn}, x_1,\ldots x_{\nnnn-1})$$

In this subsection we will frequently need to talk about $\nnnn$-tuples (of variables, or of elements of some structure), and
special attention will be paid to the first element of such tuple. For this reason, we will use an overline arrow and overline bar to denote tuples
of $\nnnn-1$ elements. We will write $ x_1,\vec x$ instead of $ x_1, x_2,\ldots x_{\nnnn}$
while by $\bar s$ we will mean a tuple $[s,s,\ldots s]$ of length $\nnnn-1$.


Now, we can use these new notations to define $\beta_s$ as:
~
$$ CYCLIQ(x_1, \vec{x}) \wedge CYCLIQ(y_1, \vec{y}) \wedge CYCLIQ(\venus, \overline{\venus}) \wedge CYCLIQ(\mars, \overline{\venus})$$

\noindent
and to define $\beta_b$ as  $CYCLIQ(x_1, \vec{x}) \wedge CYCLIQ(y_1, \vec{y}) \wedge x_1 \neq y_1 $.\medskip

The main lemma of this subsection is:

\begin{lemma}\label{osiolek}  $\beta_s$ and $\beta_b$ multiply by $ \frac{(\nnnn + 1)^2}{2\nnnn} $
\end{lemma}

The proof of Lemma \ref{osiolek} occupies the rest of this subsection.\medskip

To see that the condition ($=$) of Definition \ref{def:mnozenie} is satisfied for  $\beta_s$ and $\beta_b$,
take $D$ as
the canonical structure of the query
$CYCLIQ(\venus, \overline{\venus}) \wedge CYCLIQ(\mars, \overline{\venus})$. The active domain of this $D$ is of size 2, and
it is an easy exercise to verify that in this case
$\beta_s(D)=(\nnnn + 1)^2$ and $\beta_b(D)=2\nnnn$.

For the proof of condition ($ \leq$) from Definition \ref{def:mnozenie} consider some non-trivial
database $D$, which will remain fixed throughout this proof.
We  need to show that:
~
$$\;\;\;\;\;{\beta_b(D)} \geq \frac{2\nnnn}{(\nnnn + 1)^2}\beta_s(D)\;\;\;\;\;\;\;\;\;\;(*) $$
\medskip

At first, notice that the inequality is trivially satisfied if $\beta_s(D) = 0$.
 We may thus consider only the case when $\beta_s(D) > 0$.
 In such situation (*) is equivalent to the inequality:

$$\;\;\;\;\;\frac{\beta_b(D)}{\beta_s(D)} \geq \frac{2\nnnn}{(\nnnn + 1)^2}\;\;\;\;\;\;\;\;\;\;\;\;\;\;\;\;(**) $$
\medskip

 Before we can proceed, we need a series of  definitions:

\begin{definition}
$\bullet$~
We call a tuple $[s_1, \vec{s}]$ of elements of $D$ a {\em cyclique} if  $D\models CYCLIQ(s_1, \vec{s})$.


\noindent
$\bullet$~ For a cyclique  $[s_1, \vec{s}]$, for a natural number $0\leq k <\nnnn$, and for a tuple $[t_1, \vec{t}]$
we will say that  $[t_1, \vec{t}]$ is a cyclic $k$-shift of  $[s_1, \vec{s}]$ if
$\forall_{1 \leq i \leq n}$ $s_i = t_{((i+k) \text{ mod } n) + 1}$.

\noindent
$\bullet$~  We write  $[s_1, \vec{s}] \Bumpeq [t_1, \vec{t}]$ if there exists $k$ such that
 $[t_1, \vec{t}]$ is a cyclic $k$-shift of  $[s_1, \vec{s}]$.

\end{definition}

It is of course easy to see that
if $[s_1, \vec{s}] \Bumpeq [t_1, \vec{t}]$ then $[t_1, \vec{t}]$ is also a cyclique,
and that $\Bumpeq$ is an equivalence relation on cycliques.
 The equivalence class of  cyclique $C$ with respect to $\Bumpeq$ will be denoted as $cyclass(C)$.

Note that each cyclass has at least one and at most $\nnnn$ elements.
We are going to consider three kinds of cycliques:

\begin{definition}
\noindent
$\bullet$~
 We say that a cyclique $C$ is {\em homogeneous}, if \hspace{0.5mm}  $|cyclass(C)|=1$
 (note that it holds when $C=[s, \bar{s}]$ for some element $s$ of $D$).

\noindent
$\bullet$~
For a non-homogeneous cyclique $C$, we say that $C$ is {\em degenerate} if \hspace{0.5mm} $|cyclass(C)|<\nnnn$.

\noindent
$\bullet$~
A cyclique which is neither homogeneous nor degenerate will be called {\em normal}.
\end{definition}

Note that the above definition does not depend on the choice of the representative of the equivalence class of $\Bumpeq$,
so we can also speak about cyclasses being homogeneous, degenerate or normal. As an example of the new notions, notice
also that  $[\venus, \bar{\venus}]$ is a homogeneous cyclique and    $[\mars, \bar{\venus}]$ is a normal one.

 The proof of the next lemma, which will later be useful,
is an easy exercise in elementary group theory:

\begin{lemma}\label{grupy} If $ C$ is a degenerate cyclique  then $|cyclass(C)|\leq \nnnn/2$.
\end{lemma}

Now we are ready to come back to the proof of Lemma \ref{osiolek}.\smallskip

Inequality (**) begs for a probabilistic interpretation.
Consider an experiment in which we
draw randomly two cycliques $[s_1, \vec{s}] $ and $  [t_1, \vec{t}]$ from $D$ (with repetitions, independently and uniformly).
Denote by $ \mathrm{\mathrm{diff}}$ the  event that $s_1\neq t_1$. Then $\mathbb{P}[\mathrm{diff}] = {\beta_b(D)}/{\beta_s(D)}$, so to show that (**) is true, it will be sufficient to show that  $\mathbb{P}[\mathrm{diff}] \geq {2\nnnn}/{(\nnnn + 1)^2}$.

For two sets $X,Y$ of cycliques by $A(X,Y)$ we will denote the event that one of the two drawn cycliques comes from $X$ and another from $Y$ (in any order). Denote by $H$ the set of all homogeneous cycliques, and let $G=cyclass([\mars, \bar{\venus}])$. We know that $H$ at least contains $[\venus, \bar{\venus}]$.

The main milestone in the proof of Lemma \ref{osiolek} is:

\begin{lemma}\label{mul}
Each of the following conditions ({\bf a})-({\bf d}) implies that $\mathbb{P}[\mathrm{diff}\;|\;A(X,Y)]\geq  {2\nnnn}/{(\nnnn + 1)^2}$:\medskip

\noindent ({\bf a}) $Y$ is any cyclass and $X$ is a degenerate cyclass;

\noindent ({\bf b}) $X=Y=G\cup H$;

\noindent ({\bf c}) $X$ and $Y$ are two distinct normal cyclasses;

\noindent ({\bf d}) $X$ is a normal cyclass other than $G$ and $Y=X\cup H$.

\end{lemma}

Indeed, notice that every possible pair of cycliques falls into exactly one of the events mentioned in Lemma \ref{mul} ({\bf a})-({\bf d}).
So, once Lemma \ref{mul} is proved,  Lemma \ref{osiolek} will follow  by a trivial application of the Law of Total Probability.\medskip

\noindent
{\em Proof of Lemma \ref{mul}:}

({\bf a}) Suppose a cyclique $C=[s_1, \vec{s}]$ from $Y$ was picked and now we are about to pick a cyclique $C'$ from $X$.
Since $X$ is a degenerate cyclass (and not a homogeneous one),
there is certainly at least one  cyclique $[t_1, \vec{t}]$ in $X$ satisfying $t_1\neq s_1$. And, by Lemma \ref{grupy}, there are at most
$\nnnn/2$ cycliques in $X$. Hence, {\bf \small for each} $C\in Y$,  the probability that $[C,C']$ will be in $\mathrm{diff}$ is at least $2/\nnnn$,
which is more than  ${2\nnnn}/{(\nnnn + 1)^2}$.

({\bf b}) First consider the case that $|H|=1$, which means that  $[\venus, \bar{\venus}]$ is the only homogeneous cyclique.
Then there are $(\nnnn+1)^2$ pairs in $A(X,Y)$ and among them there are $2n$ pairs in $\mathrm{diff}$. So the  inequality from
Lemma \ref{mul} is true in this case (and, in this case, turns into the equality from condition (=)).

Now let us consider the situation when  $|H|>1$.
Suppose some $[s, \bar{s}]$  for $s\neq \venus$ was picked, from $H$, as one element of the pair.
And let us now pick  a cyclique  $C=[t_1, \vec{t}]$ as  the other element of this pair. Then, if $C\in H$ then
the probability that $t_1\neq s$ is at least $1/2$ (because there are at least 2 elements in $H$) and if
$C\in G$ then this probability is 1 (if $s\neq \mars$) or $\nnnn-1/\nnnn$ (if $s = \mars$), in both cases more than ${2\nnnn}/{(\nnnn + 1)^2}$.

({\bf c}) Suppose $X=cyclass([t_1, \vec{t}])$ and $Y=cyclass([s_1, \vec{s}])$. Now, if for each $t_i$ there exist
at least two elements in the tuple $[s_1, \vec{s}]$ which are different than $t_i$ then
$\mathbb{P}[\mathrm{diff}\;|\;A(X,Y)]\geq 2/\nnnn > {2\nnnn}/{(\nnnn + 1)^2}$ and we are done. Analogously, if for each $s_i$ there are
at least two elements of the tuple $[t_1, \vec{t}]$ which are different than $s_i$ then we are done.

So suppose there is a $t_i$ such that all elements of $[s_1, \vec{s}]$ but one are equal to $t_i$ and that there is
$s_j$ such that all elements of $[t_1, \vec{t}]$ but one are equal to $s_j$. Now, if $t_i\neq s_j$ then among the
$\nnnn^2$ possible pairs (we imagine that first we pick an element from $X$ and then from $Y$)
there are at least $(\nnnn-1)^2$ pairs in $\mathrm{diff}$, a fraction much greater than  ${2\nnnn}/{(\nnnn + 1)^2}$.
If $t_i = s_j$ then there are at least $2\nnnn-2$ pairs in $\mathrm{diff}$, and $2\nnnn-2/\nnnn^2 > {2n}/{(\nnnn + 1)^2}$.

({\bf d})  There are  $\nnnn(\nnnn + 2|H|)$ possible pairs in $A(X,Y)$.
This number includes $\nnnn^2$ pairs which do not involve a cyclique from $H$ and among them there are
at least $2\nnnn-2$ pairs in $\mathrm{diff}$ (like in the proof of ({\bf c})). There are also $\nnnn|H|$ pairs which involve a cyclique from
$H$ as the first element and $\nnnn|H|$ pairs which involve a cyclique from
$H$ as the second element.

If $|H|=1$ then among the $2\nnnn$ pairs in  $A(X,Y)$ which involve a cyclique from $H$ there are at least 2 in $\mathrm{diff}$.
So, in this case, $\mathbb{P}[\mathrm{diff}\;|\;A(X,Y)]\geq {2\nnnn}/{\nnnn(\nnnn+2)}$, which is greater than $ {2\nnnn}/{(\nnnn + 1)^2}$.

If $|H|>1$ then for every cyclique $[t_1, \vec{t}]$ from $X$ there are at least $2|H|$ pairs in $A(X,Y)$ which
involve  $[t_1, \vec{t}]$ as one of the elements and a cyclique from $H$ as the other one. At least $2(|H|-1)$ of them are in $\mathrm{diff}$.
So, in total, there are
at least $\nnnn(2(|H|-1))$ such pairs in $\mathrm{diff}$, and, in this case:
~
$$ \mathbb{P}[\mathrm{diff}\;|\;A(X,Y)]\geq \frac{2\nnnn-2+ \nnnn(2(|H|-1)) }{\nnnn(\nnnn + 2|H|)} \geq \frac{2\nnnn}{(\nnnn + 1)^2} $$

This ends the proof of  Lemma \ref{mul} and Lemma \ref{osiolek} \qed

\subsection{Constructing  $ \alpha'_s $ and $\alpha_b $: fine tuning}\label{finetuning}

The queries $\beta_s$ and $\beta_b$ from the previous subsection do a good job multiplying by arbitrarily
huge numbers.
There is a slight problem however: they only can multiply by numbers of the form $ (\nnnn+1)^2/2\nnnn$,
and  there is no way to find, for an arbitrary natural number $\ccc$ , a natural number $\nnnn$ such that $(\nnnn+1)^2/2\nnnn=\ccc$.

 Some fine tuning is needed: we will  construct queries $\gamma_s$ and $\gamma_b$ which will multiply  by something slightly less than 1,
 and then we will define $\alpha_s=\beta_s\wedgebar \gamma_s$ and $\alpha_b=\beta_b\wedgebar \gamma_b$ and use Lemma \ref{widacprzeciez}.

 Let us remark here that, for reasons which will become clear in Section \ref{dyskusja}, we are not able to multiply by anything bigger than 1, without using an inequality in the $b$-query. And we cannot afford having inequality in $\gamma_b$, because we only want to
 have one in  $\alpha_b$ and there already was one in $\beta_b$.

 But, as it turns out, multiplication by a number smaller than 1 does not require
 inequality in $\gamma_b$.

For a unary relational symbol $U$, and for some fixed new relational symbol $P$, of arity $\mmmm$ define formula
$CYCLIQ_U(x_1, x_2,\ldots x_{\mmmm})$ as:
~
$$P(x_1, x_2,\ldots x_{\mmmm})\wedge P(x_2,\ldots x_{\mmmm}, x_1) \wedge \ldots \wedge P(x_\mmmm, x_1,\ldots x_{\mmmm-1}) \wedge  U(x_1) \wedge  U(x_2)\wedge \ldots \wedge  U(x_{\mmmm})$$

Like in Section \ref{workhorse} we will be using the notations $\bar s$ (to denote a tuple of identical elements, this time of length $\mmmm-1$),
and $\vec x$ (for $x_2, x_3,\ldots x_\mmmm$).
Let now $A$ and $B$ be two new unary relation symbols.
Define $\gamma_s$ as $\gamma'_s\wedge \gamma''_s$ where:

$$\hspace{12mm}\gamma'_s = CYCLIQ_A(\mars, \bar{\venus})\wedge B(\mars)
\hspace{20mm} \gamma''_s =  CYCLIQ_B(x_1, \vec{x}) \wedge A(x_1) $$\smallskip

And define   $\gamma_b$ as $\gamma'_b\wedge \gamma''_b$ where:

$$\gamma'_b =  CYCLIQ_A(y_1, \vec{y})\wedge B(y_1)
\hspace{20mm}
\gamma''_b =  CYCLIQ_B(x_1, \vec{x})  $$\smallskip

\begin{lemma}\label{mnozenie2} The queries $\gamma_s$ and $\gamma_b$  multiply by  $\frac{\mmmm-1}{\mmmm}$.
\end{lemma}

\noindent {\em Proof:} Let us start from condition ($=$) of Definition \ref{def:mnozenie}. We need to construct a structure $D$ such that:
$${\gamma_s(D)} = \frac{\mmmm-1}{\mmmm}\gamma_b(D)\neq 0$$

So take $D$ as a disjoint union of the canonical structure of $\gamma'_s $ and of the canonical structure
of the query:
~
$$CYCLIQ_B(x_1, \vec{x}) \wedge A(x_1) \wedge A(x_2)\ldots \wedge A(x_{\mmmm-1})$$\smallskip

Now one can easily verify that $\gamma'_s(D)=1$ and $\gamma'_b(D)=1$
(notice that the last subscript in the query above is {\bf\small not} $\mmmm$ but $\mmmm-1$)
while $\gamma''_s(D)=\mmmm-1$ and $\gamma''_b(D)=\mmmm$.

Proving that condition ($\leq$) of Definition \ref{def:mnozenie} holds true for $\gamma_s$ and $\gamma_b$ is a little  bit more complicated.
Recall that we need to show that
for each non-trivial $D$ there is ${\gamma_s(D)} \leq  \frac{\mmmm-1}{\mmmm}\gamma_b(D)$. So let now $D$ be some fixed non-trivial database,
with $\gamma_s(D)\neq 0$.

Analogously to Section \ref{workhorse} a tuple satisfying
the query $CYCLIQ_U$ will be called a {\em  $U$-cyclique}. Notice that the concepts we introduced in Section \ref{workhorse} survive in
this new context:
a cyclic $k$-shift of a $U$-cyclique is again a $U$-cyclique, so one can again consider the equivalence relation $\Bumpeq$ and cyclasses.

For a unary relation $V$ we will also use the term $U$-cyclique$^V$ for such a $U$-cyclique  $[s_1,\vec{s}]$ for which $V(s_1)$ also holds.

First observation we can make is that
$\gamma''_s(D)\leq \gamma''_b(D)$: this is because $\gamma''_b(D)$ is the cardinality of the set of all $B$-cycliques in $D$, and
$\gamma''_s(D)$ is the cardinality of its subset: the set of all $B$-cycliques$^A$.

Notice also that $\gamma'_s(D)=1$, because it only mentions constants. So if $\gamma'_b(D)\geq 2$ then
${\gamma_s(D)} \leq \frac{\gamma_b(D)}{2}$ and we are done.

So, from now on we assume that   $\gamma'_b(D)=1$.
This means that there is exactly one $A$-cyclique$^B$ in $D$. And we already know one such $A$-cyclique: it is $[\mars, \bar{\venus}]$.

What remains to be proved is that  ${\gamma''_s(D)} \leq  \frac{\mmmm-1}{\mmmm}\gamma''_b(D)$.

Now, using the language of probabilities again, let ${\mathcal A } $ be the event that a randomly (uniformly) picked $B$-cyclique is a  $B$-cyclique$^A$.
We need to show that $\mathbb{P}[{\mathcal A }]\leq \frac{\mmmm-1}{\mmmm}$. To this end it will be
enough to show, for each $B$-cyclique $C$, that $\mathbb{P}[{\mathcal A }\;|\;cyclass(C)]\leq \frac{\mmmm-1}{\mmmm}$ (this is because
the cyclasses constitute a partition of the set of all $B$-cycliques). But, for each $B$-cyclique $C$ it holds that
$|cyclass(C)|\leq \mmmm$. So ($\leq$) will be proven once we show that for each $B$-cyclique $C$ there exists a
$B$-cyclique $C'$, which is not a $B$-cyclique$^A$, and such that $C'\Bumpeq C$.

So suppose, towards contradiction, that there is some $C$ such that all $B$-cycliques in $cyclass(C)$ are $B$-cycliques$^A$.
This would imply, that all elements of $C$ satisfy $A$ and, in consequence, all cycliques in $cyclass(C)$ would also be
$A$-cycliques$^B$. Recall that we assumed that there is only one $A$-cyclique$^B$ in $D$ so  $cyclass(C)$ must be a singleton,
with $C=[s, \bar{s}]$ for some $s$. But, because of the same assumption, $[s, \bar{s}]$ must be equal to $[\mars, \bar{\venus}]$,
which cannot be true in a non-trivial database.
\qed


Now take $\nnnn=2\ccc-1$ and  $\mmmm=\nnnn+1$ and put  $\alpha_s = \beta_s\wedgebar \gamma_s $ and   $\alpha_b = \beta_b\wedgebar \gamma_b $.
Notice that:
~
$$ \frac{(\nnnn+1)^2}{2\nnnn} \;  \frac {\mmmm-1}{\mmmm}=  \frac{(\nnnn+1)^2}{2\nnnn} \;  \frac{\nnnn}{\nnnn+1}=\frac{\nnnn+1}{2} =\ccc   $$

So, indeed, by Lemma \ref{widacprzeciez}, $\alpha_s$ and $\alpha_b$ multiply by $\ccc$.
This ends the proof of Theorem \ref{th-po-jednej-nier} (using Theorem \ref{th-z-c}).


\section{Proof of  Theorem \ref{th-z-c}}\label{sec:glowne-dowod-1}

In this section we prove Theorem \ref{th-z-c}, stated in Section \ref{twierdzenia}.

\subsection{The source of undecidability}\label{source-of-u}

As the source of undecidability we are going to use;

\begin{lemma}\label{source-of-und}
{\em The problem:}\medskip

Given are a natural number $\cccc \geq 2 $ and
two polynomials, $P_s$ and $P_b$, of numerical\footnote{We will call the variables ranging over {$\mathbb N$} {\em numerical variables} to
distinguish them from the first order logic variables that occur in the queries. } variables $\xxi_1, \xxi_2,\ldots  \xxi_\nnn$, with natural coefficients,
which additionally satisfy the following conditions:\medskip

\noindent
$\bullet$~
 $P_s= \Sigma_{m=1}^{\mmm} c_{s,m}{\mathbb T}_m$ and $P_b= \Sigma_{m=1}^{\mmm} c_{b,m}{\mathbb T}_m$, where ${\mathbb T}_m$, for $m\in \{1,\ldots \mmm\}$ are monomials,
and $c_{s,m}, c_{b,m}\in \mathbb N$ are coefficients;\smallskip

\noindent
$\bullet$~  there exists  $\ddd\in \mathbb N$ such that each monomial ${\mathbb T}_m$, for  $m\in \{1,\ldots \mmm\}$, is of degree exactly $\ddd$;\smallskip

\noindent
$\bullet$~ $\xxi_1$ occurs as the first variable in
each ${\mathbb T}_m$;\smallskip

\noindent
$\bullet$~  $1\leq c_{s,m}\leq c_{b,m}$ holds for each $m\in \{1,\ldots \mmm\}$.\smallskip

\noindent
Does the inequality  $\cccc P_s(\Xi(\vec \xxi)) \leq  \Xi(\xxi_1)^\ddd P_b(\Xi(\vec \xxi))$ hold for every  valuation $\Xi:\{\xxi_1, \xxi_2,\ldots  \xxi_\nnn\}\rightarrow \mathbb N$?\medskip

\noindent
{\em is undecidable}.
\medskip

\end{lemma}

Proof of Lemma \ref{source-of-und} is an  easy exercise (assuming, of course, undecidability of
the Hilbert's 10th problem),
but for the sake of completeness it will be included in the Appendix.

Notice that, in $P_s$ and $P_b$, we use the $s$ and $b$ again, and again they stand for
``small'' and  ``big''. This is how we think of the two polynomials: it is undecidable whether the ``small one'', multiplied by $\cccc $ can ever surpass
the ``big one'' (strictly speaking, the monomial $\xxi_1^\ddd$ is not part of $P_b$, but we find it convenient to imagine it is).

In order to prove Theorem 1 we will construct, for a natural number $\cccc $ and for
two polynomials, $P_s$ and $P_b$, like in Lemma \ref{source-of-und},
a tuple $[\ccc, \phi_s, \phi_b]$, consisting of a natural number and two conjunctive queries, such that
the the two conditions are equivalent:\medskip

\noindent
$\spadesuit$ There exist a valuation $\Xi:\{\xxi_1, \xxi_2,\ldots  \xxi_\nnn\}\rightarrow \mathbb N$
such that: \hspace{1mm} $\cccc P_s(\Xi(\vec \xxi)) > \Xi(\xxi_1)^\ddd P_b(\Xi(\vec \xxi))$.\medskip

\noindent
$\clubsuit$ There exists a non-trivial database $D$ such that: \hspace{1mm} $\ccc \phi_s(D)> \phi_b(D)$.\medskip

Notice that the $\ccc$ in the output tuple $[\ccc, \phi_s, \phi_b]$ of our reduction
is not the $\cccc $ in the input tuple $[\cccc , P_s, P_b]$.

\subsection{High-level definition of $\phi_s$ and $ \phi_b$ }

In Sections \ref{introducing}-\ref{serious-punish} we are going to define
queries $\Arena$, $\pi_s$, $ \pi_b$, $\zetaup_b$ and $\delta_b$.
Then $\phi_s$ will be defined as 
$\Arena \; \wedgebar \; \pi_s$ and $\phi_b$ will be defined as $\pi_b \; \wedgebar \; \zetaup_b \; \wedgebar \; \delta_b$.

In Section \ref{pieces-together} we will show that the equivalence 
$\spadesuit 
\; 
\Leftrightarrow
\; 
\clubsuit $ 
holds 
for such $\phi_s$ and $\phi_b$, and for 
the number $\ccc$ as defined in Section \ref{punish-slight}.


\subsection{Introducing the queries $\pi_s$ and $ \pi_b$.  }\label{introducing}

We assume that polynomials $P_s$ and $P_b$ are like in Lemma \ref{source-of-und}, so in particular they
have the same monomials ${\mathbb T}_1, {\mathbb T}_2, \ldots {\mathbb T}_{\mmm}$, all of them of degree $\ddd$. Recall that
coefficients before ${\mathbb T}_1, {\mathbb T}_2, \ldots {\mathbb T}_{\mmm}$ in $P_s$ are (respectively) $c_{s,1}, c_{s,2}, \ldots c_{s,\mmm}$ and the
coefficients before ${\mathbb T}_1, {\mathbb T}_2, \ldots {\mathbb T}_{\mmm}$ in $P_b$ are (respectively) $c_{b,1}, c_{b,2}, \ldots c_{b,\mmm}$.

Recall also that for each $1\leq m\leq \mmm$ we have $1\leq c_{s,m} \leq c_{b,m}$.

Notice that we use $m$ as a natural number ranging from $1$ to $\mmm$. In a similar manner we will always
use $n$ as a natural number ranging from $1$ to $\nnn$ (recall that $\nnn$ is the number of variables in $P_s$ and $P_b$) and $d$ as a natural number ranging from $1$ to $\ddd$. Also, define $\lll=\mmm+\nnn+2$.

Let $\Sigma_0$ be the schema comprising a binary relation $S_m$ for each $ m\in  \{1,\ldots \mmm\} $  (so that we have one relation $S$ for each
monomial in $P_s$ and $P_b$),
a binary relation $R_d$ for each $d \in \{1,\ldots \ddd\}$,
and a binary relation $E$. And let $\Sigma= \Sigma_0\cup \{X\} $ for some binary relation $X$.\smallskip

Now, define the query $\pi_s$ as:

$$ \bigwedge_{m \in \{1,\ldots \mmm\} } S_{m}(x,x)\wedge S_{m}(x,x^m_{c_{s,m}})
\;\;\wedge\;\bigwedge_{m \in \{1,\ldots \mmm\} } \bigwedge_{ 1\leq k < c_{s,m}} S_{m}(x^m_{k+1},x^m_{k})
\;\;\wedge\; \bigwedge_{d \in \{1,\ldots \ddd\} } R_d(x,y_d)\wedge X(y_d,z_d)$$\medskip

and let $\pi_b$ be:\bigskip

\noindent
$$\bigwedge_{m \in \{1,\ldots \mmm\} }\; S_{m}(x,x)\wedge S_{m}(x,x^m_{c_{b,m}})
\;\wedge\;\bigwedge_{m \in \{1,\ldots \mmm\} }\;\bigwedge_{ 1\leq k < c_{b,m}}\; S_{m}(x^m_{k+1},x^m_{k})
\;\wedge\;\bigwedge_{d \in \{1,\ldots \ddd\} }\; R_d(x,y_d)\wedge X(y_d,z_d)$$
$$  \hspace{94mm} \wedge\;\bigwedge_{d \in \{1,\ldots \ddd\} }\; R_1(x,y'_d)\wedge X(y'_d,z'_d)    $$\medskip

It will be helpful at this point to get some understanding of the structure of  $\pi_b$: the (canonical structure of the) query is a star, with $x$ as its center.
For each monomial ${\mathbb T}_m$ in $P_b$ there is an ``$S_m$-ray'' in this star (radiating from the center, as the rays normally do) of length equal to
the coefficient before ${\mathbb T}_m$ in $P_b$. There is also an $S_m$-loop in $x$ which is very important, as you are soon going to see.

Apart from the ``$S_m$-rays'' there are also  rays of length two,
$2\ddd$ of them, as many  as there are numerical variable occurrences in each term of 
$\xxi_1^\ddd P_b$. Each such ray begins with $R_d$ (this $d$ indicates which numerical variable we have in mind) and then comes the $X$ which
is supposed (as you are going to see in Section \ref{howcompute}) to represent the valuation $\Xi$. The fact that all the 
``rays'' representing $\xxi_1^\ddd$ begin from $R_1$ is related to the third  condition from Lemma \ref{source-of-und}.

The structure of $\pi_s$ relates to $P_s$ as $\pi_b$ relates to $P_b$.

\begin{lemma}\label{trik-ktorego-oni-nie-widzieli}
For every database $D$ there is $\pi_s(D)\leq\pi_b(D)$.
\end{lemma}

\noindent {\em Proof:~} Let us start from a simple observation. Suppose for some queries  $\rho_b$ and $\rho_s$  there exists an
{\em onto} mapping $h$ from 
the variables of $\rho_b$ to the variables of $\rho_s$ which is a homomorphism of queries. Then for every database $D$ there is $\rho_s(D)\leq\rho_b(D)$.

Indeed, if $h$ is a mapping as in the previous paragraph, then the function $H$ (of argument $g$) defined as $H(g)= g\circ h$ is a 1-1 function from $Hom(\rho_s,D)$ to $Hom(\rho_b,D)$. Obviously, if
$g\in Hom(\rho_s,D)$ then $g\circ h \in Hom(\rho_b,D)$. To see that $H$ is 1-1, take $g, g'\in Hom(\rho_s,D)$  such that $g\neq g'$. Then there must be
$v_s$, a variable in $\rho_s$, such that $g(v_s)\neq g'(v_s)$. Let $v_b$ be such a variable of $\rho_b$ that $h(v_b)=v_s$ (recall that $h$ is onto).
Then $gh(v_b)\neq g'h(v_b)$.

What remains to be shown is that there exists an {\em onto} homomorphism $h$ from $\pi_b$ to $\pi_s$. 

To this end, first of all notice that $Var(\pi_s)\subseteq Var(\pi_b)$. Let us first define $h$ on the variables which are 
both in $Var(\pi_s)$ and in $ Var(\pi_b)$ as the identity:

\begin{center}
$h\upharpoonright Var(\pi_s) = id_{Var(\pi_s)}$
\end{center}
At this point we can already be sure that $h$ will be {\em onto}. Now we need to define its values for the variables in
 $Var(\pi_b)\setminus Var(\pi_s)$  in such a way that the resulting $h$ is indeed a homomorphism. 
 
 For each variable $x^m_k\in Var(\pi_b)\setminus Var(\pi_s)$  define $h(x^m_k)=x$. See how, thanks to the
 $S_m$-loops at $x$, the $S_m$-rays in $\pi_b$ are now homomorphically mapped onto the $S_m$-rays in $\pi_s$ (this is the only 
 place in the entire paper where we use the condition, from Lemma \ref{source-of-und}, 
 that the coefficients in $P_b$ are at least equal to the respective
 coefficients in $P_s$).
 
 What still remains to be defined are the values of $h$ for the variables (other than $x$) in the subquery $\bigwedge_{d \in \{1,\ldots \ddd\} }\; R_1(x,y'_d)\wedge X(y'_d,z'_d)    $.
 It is not hard to guess that $h$ will map all the $y$'s to $y_1$ and it will map all the $z$'s to $z_1$. \qed

\subsection{$\Arena$, and how $\pi_s$ and $ \pi_b$ compute $P_s$ and $P_b$.}\label{howcompute}

The query $\Arena=\Arena_\pi\wedge\Arena_\delta$ over $\Sigma_0$ 
will be defined as a conjunction of facts mentioning only constants (and no variables). Because of that,
for every database $D$ we will have $\Arena(D)\in\{0,1\}$:
the value is $1$ if all the atomic formulas in $\Arena$ are indeed facts in $D$ and it is $0$ otherwise.

Definition of the query $\Arena_\pi$ will follow now
 and $\Arena_\delta$ will be defined in Section \ref{serious-punish}, together with the query $\delta_b$.

Let ${\mathcal P}\subseteq \{1,\dots \nnn\}\times \{1, \dots \ddd\} \times \{1,\dots \mmm\}$ be the relation saying
 which numerical variable constitutes which argument in which of the monomials:
${\mathcal P}(n,d,m)$ means that $\xxi_n$ is the $d$-th variable\footnote{If  variable $\xxi_n$ occurs in  ${\mathbb T}_m$  more than once then
${\mathcal P}(n,d,m)$  will be true for more than one number $d$.} in ${\mathbb T}_m$.

Now define $\Arena_\pi$  as the  query:

$$\bigwedge_{[n,d,m]\in {\mathcal P}}\; R_d(a_m,b_n) \; \wedge \;
\bigwedge_{m,m' \in \{1,\ldots \mmm\} }\; S_{m'}(a_m,a_m) \;\;\wedge \;\;
\bigwedge_{m \in \{1,\ldots \mmm\} }\; S_{m}(a_m,a) \wedge  S_{m}(a, a) $$\medskip

Notice that we have one constant ($a_m$) for each monomial and one constant ($b_n$) for each numerical variable.
As we said, $\Arena_\delta$ is not going to be defined right now. All you need to know at this point is
that the only relation  $\Arena_\delta$  mentions is $E$, which does not appear in $\Arena_\pi$.

Let $D_{\Arena}$ be the canonical structure of the query $\Arena$.

\begin{definition}\label{klasyfikacja-baz}
A database $D$ over $\Sigma$, such that $D\models \Arena$, will be called:

\noindent
$\bullet$~ {\em correct} if $D\upharpoonright \Sigma_0 = D_{{\footnotesize \Arena}}$, where by $D\upharpoonright\Sigma_0$ we mean the
database resulting from $D$ by removing from it all atoms of the relation $X$;

\noindent
$\bullet$~ {\em slightly incorrect} if it is not correct but $D\upharpoonright \Sigma_0 \supseteq D_{\Arena}$, where $\supseteq$ is understood to be inclusion of
relational structures;

\noindent
$\bullet$~ {\em seriously incorrect} if it is neither correct nor slightly incorrect.
\end{definition}

In other words, a correct database is just $D_{\Arena}$ with some additional atoms of relation $X$,
a slightly incorrect database is $D_{\Arena}$ with some additional atoms of relation $X$ and possibly also of the relations from $\Sigma_0$,
and a seriously incorrect database is one which satisfies $\Arena$ (that is, contains a homomorphic image of $D_{\Arena}$),
but which identifies some elements of $D_{\Arena}$ (that is, this homomorphism is not 1-1).

As we already mentioned, the relation $X$ will represent a valuation of numerical variables:

\begin{definition}
For a database $D$, over the schema $\Sigma$, such that $D\models \Arena$ let us define
a valuation $\Xi_D: \{\xxi_1, \xxi_2,\ldots  \xxi_n\}\rightarrow \mathbb N$ in the following way:
~
$$ \Xi_D(\xxi_i)= (\exists x\; X(b_i,x))(D) $$

\end{definition}

Translating it to human language, notice that $ \exists x\; X(b_i,x) $ is a
boolean query, so -- when applied to $D$ -- it returns a number, which is the number of $X$-edges that begin at $b_i$.

Proof of the following lemma can be found in Appendix A:

\begin{lemma}\label{dobre-liczenie}
If $D$ is a correct database, then
$\pi_s(D)=P_s(\Xi_D(\vec \xxi))$ and $ \pi_b(D)=\Xi_D(\xxi_1)^\ddd P_b(\Xi_D(\vec \xxi))$.
\end{lemma}

Clearly, for every valuation $\Xi: \{\xxi_1, \xxi_2,\ldots  \xxi_\nnn\}\rightarrow \mathbb N$, there exists
a correct database $D$ such that $\Xi=\Xi_D$. This, together with Lemma \ref{dobre-liczenie},
implies:

\begin{lemma}\label{rownowaznosc-dla-correct}
The following two conditions are equivalent:

 $\bullet$~ there exists  a valuation $\Xi:\{\xxi_1, \xxi_2,\ldots  \xxi_\nnn\}\rightarrow \mathbb N$
such that
$\cccc P_s(\Xi(\vec \xxi)) > \Xi(\xxi_1)^\ddd P_b(\Xi(\vec \xxi))$

$\bullet$~ there exists a  correct database $D$ such that $\cccc \pi_s(D) > \pi_b(D)$.

\end{lemma}

\subsection{How to punish for slight incorrectness}\label{punish-slight}
Lemma \ref{rownowaznosc-dla-correct} almost looks like the $\spadesuit \; \Leftrightarrow \; \clubsuit$ equivalence which we are trying to prove. Almost, 
because it only works for correct databases. And we have no idea how
the values of $\pi_s(D)$ and $\pi_b(D)$ would behave if we were given
a slightly incorrect database, or a seriously incorrect one. In Sections \ref{punish-slight} and  \ref{serious-punish}, we create tools for ``punishing'' such databases.
The tools will be then used in Section \ref{pieces-together}.

Let $\Sigma_{RS}= \{S_1,\ldots S_\mmm, R_1, \ldots R_\ddd \}$. For
a relation symbol $P\in \Sigma_{RS}$, let $\jjjj^P$
denote the number of atoms of the relation $P$ in $\Arena$.

Let $\jjjj=\max(\{\jjjj^P : P\in \Sigma_{RS}\}) $ and
let $\kkkk$ be
the smallest natural number such that $(\frac{\jjjj+1}{\jjjj})^{\kkkk}\geq \cccc  $. Clearly, we also have
$(\frac{\jjjj^P+1}{\jjjj^P_m})^{\kkkk}\geq \cccc  $  for
each $P \in \Sigma_{RS} $.

 Now, again for a relation symbol $P\in \Sigma_{RS}$, let  $\zetaup^P$ be the query:
~
$$  P(w,v) \; \uparrow \; \kkkk $$
~
And let $\zetaup_b$ be defined as
$\bar\bigwedge_{P\in \Sigma_{RS}} \zetaup^P $.
Finally, define $\ccc_1 = \zetaup_b( D_{\Arena})$ and $\ccc=\cccc \ccc_1$ (recall that $\cccc $ comes from our input tuple $[\cccc , P_s, P_b]$,
and $\ccc$ is a part of the output tuple $[\ccc, \phi_s, \phi_b]$).
 Then of course:
 
 \begin{lemma} \label{ccc1-dla-correct}
 \begin{itemize}
 \item
 If $D$ is  correct then    $\zetaup_b(D) = \ccc_1 $.
 \item
 If $D\models \Arena$  then    $\zetaup_b(D) \geq 1$.
 \end{itemize}
 \end{lemma}
 
 The second claim of the above lemma follows since, in order for $\Arena$ to be satisfied in $D$, there must be
 at least one atom in $D$ of each of the relations in $\Sigma_{RS}$.
 Finally:

\begin{lemma}\label{ccc1-dla-sl-incorrect}
If $D$ is  slightly incorrect then    $\zetaup_b(D)\geq \ccc $
\end{lemma}

\noindent {\em Proof:} Clearly, for any database $D$ we have:
~
$$\;\;\;\;\;\;\;  \;\;\;\;
\zetaup_b(D)=\Pi_{P\in \Sigma_{RS} } \;\; \zetaup^P(D)
\;\;\;\;\;\;\;  \;\;
(*)  $$ 

\noindent
If $D$ is slightly incorrect then, by definition of slight incorrectness:

(a) for each $P\in \Sigma_{RS}$ there are at least as many atoms of $P$ in $D$ as in $D_\Arena$;

(b) there is $P_0\in \Sigma_{RS}$ such that there are more atoms of $P_0$ in $D$ then in $D_\Arena$.

From (a) we get that  $\zetaup^P(D) \geq  \zetaup^P(D_\Arena) $ for each $P\in \Sigma_{RS}$.
From (b) we get that there are at least $\jjjj^{P_0}+1$ atoms of the relation $P_0$ in $D$. It implies
that \hspace{2mm}
${\zetaup^{P_0}(D)}/{\zetaup^{P_0}(D_\Arena)}\geq \cccc $.
So from (*), from (a) and from (b) we get that  \hspace{2mm}
$\zetaup_b(D)\geq \cccc \zetaup_b(D_\Arena) = \ccc$.
\qed
%
%
%
%
%
%
%
%

\subsection{How to punish for serious incorrectness} \label{serious-punish}

Recall that the query $\Arena_\pi$ mentions constants $a, a_1, a_2,\ldots a_\mmm$ and $b_1, b_2,\ldots b_\nnn$ 
and that $E\in  \Sigma_0$ is a binary relation symbol which was not used so far.
It is now time to define   $\Arena_\delta$, as the query:\medskip

 $E(\mars,\mars)\wedge E(\venus, a)\wedge E(a, a_1)\wedge E(a_1, a_2)\wedge \ldots \wedge E(a_{\mmm-1}, a_\mmm) \wedge $
 
 $ E(a_\mmm, b_1) \wedge  E(b_1,b_2) \wedge \ldots \wedge E(b_{\nnn-1}, b_\nnn)\wedge E(b_\nnn, \venus)$
 \medskip
 
In words, $\Arena_\delta$ comprises the self-loop $E(\mars,\mars)$ and an $E$-cycle, of length $\lll$ 
(recall that $\lll=\nnn+\mmm+2$) containing $\venus$ and all the constants from $Arena_\pi$.

For a natural number $l$ let $\delta_{b,l}$ be the query:
$$ E(z_{1},z_{2}) \wedge E(z_{2},z_{3})\wedge\ldots   E(z_{l-1},z_{l})\wedge E(z_{l},z_{1}) $$

Let $L=\{1,2,\ldots, \lll-1\}\cup \{\lll+1\}$. Define  $\delta_b$ as $(\bar\bigwedge_{ l\in L} \; \delta_{b,l})\; \uparrow \; \ccc$.

In human language the meaning of $\delta_b(D)$ is as follows: for each $l\in L$ count all (the homomorphic images of) 
cycles of length $l$ in $D$, take a product of all the 
$|L|$ numbers you got and raise this product to the power $\ccc$.

\begin{lemma}\label{petelka1+}
Suppose $D$ is  such that $D\models \Arena$. Then $\delta_b(D)\geq 1$.
\end{lemma}

\noindent {\em Proof:~} Recall that if $D\models \Arena$ then $E(\mars,\mars)$ must be true in $D$. Then $\delta_b$ can be satisfied in $D$ by
mapping all its variables to $\mars$.\qed

\begin{lemma}\label{petelka1}
 If $D$ is a correct database then $\delta_b(D)= 1$.
\end{lemma}

\noindent {\em Proof:~}  $\Arena$, and hence also any other correct database,
contains two cycles (with respect to the relation $E$): one of length 1 (namely, $E(\mars,\mars)$) 
and one of length $\lll$. 

But $L$ contains all natural numbers up to $\lll+1 $ {\em except for} $\lll$. So, for every $l\in L$,
there is exactly one (homomorphic image of a) cycle of length $l$ in $\Arena$ -- the loop $E(\mars,\mars)$. 
In consequence  $\delta_b(D)$ is the product of $|L|$ ones, raised to power $\lll$, and this equals 1. \qed

\begin{lemma}\label{megazlepianie}
Suppose $D$ is a seriously incorrect  non-trivial database,  such that $D\models \Arena$. Then $\delta_b(D)\geq 2^\ccc \geq\ccc$.
\end{lemma}

\noindent {\em Proof:~} Suppose $D$ is a seriously incorrect non-trivial database ,  such that $D\models \Arena$.
We need to prove that:
~
$$ (\bigwedge_{ l\in L} \delta_{b,l})(D)\; \geq \; 2 $$
~
Using the argument from the proof of Lemma \ref{petelka1+} we get that for each $l\in L$ there is
$ \delta_{b,l}(D) \geq 1 $.
So what remains to be proved is that  there
exists $l\in L$ such that
$ \delta_{b,l}(D) \geq 2 $. Or, in other words, that there exists $l\in L$ and a homomorphism $h:\delta_{b,l} \rightarrow D$
which {\bf does not} map all the variables of $\delta_{b,l}$ to $\mars$. Call such homomorphisms {\em non-trivial}.

There are two cases, depending on which constants of $\Arena$ are identified in the (seriously incorrect) $D$.\smallskip

{\em Case 1. $D$ identifies $\mars$ with some  other constant of $\Arena$.} Then there indeed exists a 
non-trivial homomorphism $h\in Hom(\delta_{b,\lll+1}, D)$. To see it first notice that 
$D\models \Arena_\delta$, so there must be an $E$-cycle of length $\lll$ in $D$. One of the elements of this cycle is
identified with $\mars$, and $D\models E(\mars, \mars)$, so we have a cycle of length $\lll$  whose one vertex has a self-loop, 
and we can of course homomorphically embed $\delta_{b,\lll+1}$ (which itself is a cycle of length $\lll+1$) in such a cycle. 
But why are we sure that this homomorphism is non-trivial? This is because $D$ is non-trivial, so we are sure that $\mars$ and $\venus$ are not identified in $D$.\smallskip

{\em Case 2. $D$ does not identify $\mars$ with any other constant of $\Arena$ but it does identify some other two constants.} Then
some two vertices of the length-$\lll$ $E$-cycle from $\Arena_\delta$ are identified in $D$, which produces a shorter
cycle. In consequence, there  exists a 
non-trivial homomorphism $h\in Hom(\delta_{b,l}, D)$ for some $l<\lll$.
\qed

\subsection{Putting all pieces together} \label{pieces-together}

What remains to be shown is that  the equivalence $\spadesuit \; \Leftrightarrow \; \clubsuit$ from Section  \ref{source-of-und} holds
for $\phi_s= \Arena \wedgebar \pi_s$, for $\phi_b=\pi_b \wedgebar \zetaup_b\wedgebar \delta_b$, and for the
 $\ccc$ as in Section \ref{punish-slight}.\medskip
 
 To show that $\spadesuit \; \Rightarrow \; \clubsuit $, suppose
  $\Xi:\{\xxi_1, \xxi_2,\ldots  \xxi_\nnn\}\rightarrow \mathbb N$
is such that
 $\cccc P_s(\Xi(\vec \xxi)) > \Xi(\xxi_1)^\ddd P_b(\Xi(\vec \xxi))$.
 Take a correct database $D$ such that $\Xi_D=\Xi$. By Lemma \ref{dobre-liczenie}
we get that $\cccc \pi_s(D)>\pi_b(D)$. 
Recall that $\Arena(D)=1$.  From Lemma \ref{petelka1}  also $\delta_b(D)=1$. So
$\cccc  (\Arena \wedgebar \pi_s)(D)    >  (\pi_b \wedgebar \delta_b)(D) $.
 We also know, from Lemma \ref{ccc1-dla-correct}, that $ \zetaup_b( D)=\ccc_1$.
Multiplying both sides by $\ccc_1$, we get:
 $$ \cccc \ccc_1 (\Arena \wedgebar \pi_s)(D)    >  (\pi_b \wedgebar \zetaup_b \wedgebar \delta_b)(D) $$

Now just recall that $\ccc=\cccc \ccc_1$.\medskip

For the $\spadesuit \; \Leftarrow \; \clubsuit $ direction, suppose $\cccc P_s(\Xi(\vec \xxi)) \leq \Xi(\xxi_1)^\ddd P_b(\Xi(\vec \xxi))$ for
all valuations $\Xi$. We want to show that in such case for every non-trivial $D$ there is $\ccc \phi_s(D) \leq \phi_b(D)$.

So fix a non-trivial database $D$. If $D\not\models \Arena$ then $D\not\models \phi_s$ and there is nothing to prove.
If $D\models \Arena$ then there are
three cases:  $D$ may be correct, or  slightly incorrect, or  seriously incorrect.

If $D$ is correct then just reuse the above argument for  $\spadesuit \; \Rightarrow \; \clubsuit $.
For the remaining two cases, first recall Lemma \ref{trik-ktorego-oni-nie-widzieli}: whatever $D$ is, we know for sure that $\pi_s(D) \leq \pi_b(D)$.
We need to show that $\ccc \leq (\zetaup_b \wedgebar \delta_b)(D)$.

Now, if $D$ is slightly incorrect, then  $\delta_b(D)\geq 1$ (by Lemma \ref{petelka1+}) and $\zetaup_b (D)\geq \ccc$ (by Lemma \ref{ccc1-dla-sl-incorrect}).

If $D$ is seriously incorrect, then  $\delta_b(D)\geq \ccc$ (by Lemma \ref{megazlepianie}) and   $\zetaup_b (D)\geq 1$ (by the second claim of Lemma \ref{ccc1-dla-correct}).

This ends the proof of Theorem \ref{th-z-c}.

\section{Proof of Theorem \ref{dodatkowe}}\label{dyskusja}

\label{dyskusja}

In this Section we prove Theorem \ref{dodatkowe}. But first we need to introduce the tools.

\subsection{Operations on structures } We are
going to construct new structures (from some given structures) using two standard operations on graphs (which also apply to
any other relational structures): graph product and the the blow-up operation.
 Let us recall their definitions (see for example \cite{lovaszbook} Chapter 3.3).\medskip

\noindent
{\bf The blow-up operation.} For a relational structure $D$, with the set of vertices $V$, and for a natural number $\kkkk$, the structure $blowup(D,\kkkk)$ is defined as follows.
The set of vertices  of  $blowup(D,\kkkk)$ is  $V_\kkkk=\{[s,i]: s\in V \wedge i\in\{1,2,\ldots \kkkk\} $. Then,
for each relation\footnote{In order to keep the notation light (and avoid superscripts) we imagine that $R$ is binary, but the definition is analogous for relations of any arity. } symbol $R$,
and for each $[s,i],[r,j]\in V_\kkkk$,
the atom $R([s,i], [r,j])$ is in $blowup(D,\kkkk)$ if and only if  $R(s,r)$ is in $D$.\medskip

\noindent
{\bf The product operation.} For two structures $D_1$ and $D_2$, with sets of vertices $V_1$ and $V_2$ respectively,
their product $D_1\times D_2$
is defined as a structure whose set of vertices is $V_1\times V_2$, such that $R([s,s'],[r,r'])$ is an atom\footnote{The remark from the previous footnote applies here accordingly.} of $D_1\times D_2$ if and only if $R(s,r)$ is an atom of $D_1$ and $R(s',r')$ is an atom of $D_2$.
For a structure $D$ and natural number $\kkkk$ by $D^{\times \kkkk}$ we denote the product of $\kkkk$ copies of $D$.\medskip

\noindent
The well-known and important lemma (and also easy to prove) is:

\begin{lemma}\label{peracje}
Suppose $D$ is a structure, $\phi$ is a CQ, without inequalities,  with $\jjjj$ variables and  $\kkkk$ is a natural number. Then:

\begin{itemize}
\item[(i)] $\phi(blowup(D,\kkkk))=  \kkkk^\jjjj\phi(D) $

\item[(ii)]  $\phi(D^{\times \kkkk}))=  (\phi(D))^\kkkk $

\end{itemize}

\end{lemma}

Notice that it follows directly from Lemma \ref{peracje} (ii) that if $\varrho_s$ and $\varrho_b$ (like in Definition \ref{def:mnozenie}) are CQs without inequality then they cannot
multiply by a number greater than 1.

\subsection{Proof of Theorem \ref{dodatkowe}}

Now, Theorem \ref{dodatkowe} will follow directly from:

\begin{lemma}\label{niepodzianka}
Let $ \psi_s $ and $\psi_b $ be conjunctive queries, $\psi_b$ without inequalities and $\psi_s$ with inequalities.
Let $ \psi'_s $ be  $ \psi_s $ with all the inequalities removed. Then the two conditions are equivalent:

\noindent
(a) there exist a structure $D$, such that $\psi_s(D)>\psi_b(D)$;\\
(b) there exist a structure $D_0$, such that $\psi'_s(D_0)>\psi_b(D_0)$;

\end{lemma}

 {\em Proof of Lemma \ref{niepodzianka}:} To keep the presentation simple we will show the lemma for $\psi_s$ having exactly
 one inequality $x\neq x'$. Proof in the general case is similar (we will comment on it at the end of this Section).

 Of course, for each $D$, there is $\psi'_s(D) \geq \psi_s(D)$ so (a) implies (b).

 Now suppose there exists a structure $D_0$, such that $\psi'_s(D_0)>\psi_b(D_0)$. We are going to construct $D$
 such that $\psi_s(D)>\psi_b(D)$. One lemma we will need is:

 \begin{lemma}\label{malusienki}
 For every structure $D$ it holds that: $\psi_s(blowup(D,2))\geq \frac{\psi'_s(blowup(D,2))}{2}$\medskip
 \end{lemma}

 \noindent
 {\em Proof of Lemma \ref{malusienki}:} \hspace{1mm} It is  clear that:
 $$Hom(\psi_s,blowup(D,2)) \subseteq Hom(\psi'_s,blowup(D,2)) $$
 In order to prove Lemma \ref{malusienki} it will be enough to show that:
 $$ |Hom(\psi'_s,blowup(D,2)) \;\setminus \; Hom(\psi_s,blowup(D,2))|\leq  |Hom(\psi_s,blowup(D,2))|  $$

Translating the above into the human language, we want to show that for each homomorphism $h\in Hom(\psi'_s,blowup(D,2))$ which {\bf does not}
 map $x$ and $x'$ to different vertices of $blowup(D,2)$ there is another such homomorphism which {\bf does}.

 To this end it will be enough to construct an injection $F$ from the set
  $Hom(\psi'_s,blowup(D,2))\setminus Hom(\psi_s,blowup(D,2)) $ to the set $Hom(\psi_s,blowup(D,2))$. In other words,
  it will be enough to construct a 1-1 function $F$ which will produce,
  for each mapping (from the set of variables
  of $\psi_s$ to $blowup(D,2)$) satisfying $\psi'_s$ but not $\psi_s$
   a mapping (from the set of variables
  of $\psi_s$ to $blowup(D,2)$) which will satisfy  $\psi_s$.

So take a mapping $h$ which is in $ Hom(\psi'_s,blowup(D,2))$ but not in $ Hom(\psi_s,blowup(D,2)) $. Clearly,
  $h(x)=h(x')=[s,\epsilon]$ for some $s\in V_D$ and some $\epsilon\in\{1,2\}$.
  And define $F(h)$ as the mapping:

 $$
    (F(h))(y)=
    \begin{cases}
     [s,\epsilon]  & \text{if}\ y=x \\
      [s,3-\epsilon]    & \text{if}\ y=x' \\
     h(y)  & \text{if}\ y\neq x \text{ and if}\ y\neq x'
    \end{cases}
  $$

 It is now trivial to see that the $F$ we have just defined is indeed as required, which ends the proof of Lemma \ref{malusienki} \qed

 Let now $\jjjj$ be the number of variables in $\psi_b$. Since $\psi'_s(D_0)>\psi_b(D_0)$, and by Lemma \ref{peracje} (ii), there exists
 $\kkkk$  such that:
~
 $$  \psi'_s(D_0 ^{\times \kkkk})> 2^{\jjjj+1}\psi_b(D_0 ^{\times \kkkk})    $$\smallskip

 Now take $D=blowup(D_0 ^{\times \kkkk} ,2)$. By Lemma \ref{peracje} (i) we have:
~
 $$ \psi_b(D)=  2^\jjjj\psi_b( D_0^{\times \kkkk} )  $$

 It then follows from the choice of $\kkkk$ that:
~
 $$ \psi'_s(D) > 2\psi_b(D) $$

 Finally, from Lemma \ref{malusienki} we get that:
~
 $$ \psi_s(D) > \psi_b(D) $$

 Which ends the proof of Lemma \ref{niepodzianka} and of Theorem \ref{dodatkowe} \qed

 If there were \nnnn~ inequalities in $\psi_s$  rather than one, then we would use 2\nnnn~ rather than 2 in Lemma \ref{malusienki} and
 we would need to modify the $\kkkk$ above accordingly.

\newpage


\bibliography{references}

\begin{thebibliography}{10}

\bibitem{CV93}
S.~Chaudhuri and M.~Y. Vardi, ``Optimization of real conjunctive queries,'' in
  {\em Proceedings of the Twelfth ACM SIGACT-SIGMOD-SIGART Symposium on
  Principles of Database Systems}, PODS '93, (New York, NY, USA), p.~59–70,
  Association for Computing Machinery, 1993.

\bibitem{CM77}
A.~K. Chandra and P.~M. Merlin, ``Optimal implementation of conjunctive queries
  in relational data bases,'' in {\em Proceedings of the 9th Annual {ACM}
  Symposium on Theory of Computing, May 4-6, 1977, Boulder, Colorado, {USA}},
  pp.~77--90, {ACM}, 1977.

\bibitem{SY80}
Y.~Sagiv and M.~Yannakakis, ``Equivalences among relational expressions with
  the union and difference operators,'' {\em J. {ACM}}, vol.~27, no.~4,
  pp.~633--655, 1980.

\bibitem{K88}
A.~C. Klug, ``On conjunctive queries containing inequalities,'' {\em J. {ACM}},
  vol.~35, no.~1, pp.~146--160, 1988.

\bibitem{M97}
R.~van~der Meyden, ``The complexity of querying indefinite data about linearly
  ordered domains,'' {\em Journal of Computer and System Sciences}, vol.~54,
  no.~1, pp.~113--135, 1997.

\bibitem{CW91}
Y.~E. Ioannidis and E.~Wong, ``Towards an algebraic theory of recursion,'' {\em
  Journal of the ACM (JACM)}, vol.~38, no.~2, pp.~329--381, 1991.

\bibitem{ADG10}
F.~N. Afrati, M.~Damigos, and M.~Gergatsoulis, ``Query containment under bag
  and bag-set semantics,'' {\em Information Processing Letters}, vol.~110,
  no.~10, pp.~360--369, 2010.

\bibitem{KRS12}
E.~V. Kostylev, J.~L. Reutter, and A.~Z. Salamon, ``Classification of
  annotation semirings over query containment,'' in {\em Proc. of the 31st
  {ACM} {SIGMOD-SIGACT-SIGART} Symposium on Principles of Database Systems,
  {PODS} 2012, Scottsdale, AZ, USA, May 20-24, 2012}, pp.~237--248, {ACM},
  2012.

\bibitem{cohen09}
S.~Cohen, ``Equivalence of queries that are sensitive to multiplicities,'' {\em
  {VLDB} J.}, vol.~18, no.~3, pp.~765--785, 2009.

\bibitem{chirkova12}
R.~Chirkova, ``Equivalence and minimization of conjunctive queries under
  combined semantics,'' in {\em 15th Int. Conf. on Database Theory, {ICDT} '12,
  Berlin, Germany, March 26-29, 2012}, pp.~262--273, {ACM}, 2012.

\bibitem{KM19}
G.~Konstantinidis and F.~Mogavero, ``Attacking diophantus: Solving a special
  case of bag containment,'' in {\em Proc. of the 38th {ACM}
  {SIGMOD-SIGACT-SIGAI} Symposium on Principles of Database Systems, {PODS}
  2019, Amsterdam, The Netherlands, June 30 - July 5, 2019}, pp.~399--413,
  {ACM}, 2019.

\bibitem{KoR11}
S.~Kopparty and B.~Rossman, ``The homomorphism domination exponent,'' {\em Eur.
  J. Comb.}, vol.~32, no.~7, pp.~1097--1114, 2011.

\bibitem{AKNS20}
M.~Abo~Khamis, P.~G. Kolaitis, H.~Q. Ngo, and D.~Suciu, ``Bag query containment
  and information theory,'' in {\em Proceedings of the 39th ACM
  SIGMOD-SIGACT-SIGAI Symposium on Principles of Database Systems}, PODS'20,
  (New York, NY, USA), p.~95–112, Association for Computing Machinery, 2020.

\bibitem{IR95}
Y.~E. Ioannidis and R.~Ramakrishnan, ``Containment of conjunctive queries:
  Beyond relations as sets,'' {\em ACM Trans. on Database Systems (TODS}, 1995.

\bibitem{JKV06}
T.~S. Jayram, P.~G. Kolaitis, and E.~Vee, ``The containment problem for
  <bi>real</bi> conjunctive queries with inequalities,'' in {\em Proceedings of
  the Twenty-Fifth ACM SIGMOD-SIGACT-SIGART Symposium on Principles of Database
  Systems}, PODS '06, (New York, NY, USA), p.~80–89, Association for
  Computing Machinery, 2006.

\bibitem{KMO22}
J.~Kwiecien, J.~Marcinkowski, and P.~Ostropolski{-}Nalewaja, ``Determinacy of
  real conjunctive queries. the boolean case,'' in {\em {PODS} '22:
  International Conference on Management of Data, Philadelphia, PA, USA, June
  12 - 17, 2022} (L.~Libkin and P.~Barcel{\'{o}}, eds.), pp.~347--358, {ACM},
  2022.

\bibitem{lovaszbook}
L.~Lov{\'{a}}sz, {\em Large Networks and Graph Limits}, vol.~60 of {\em
  Colloquium Publications}.
\newblock American Mathematical Society, 2012.

\bibitem{Davis77}
M.~Davis, ``Unsolvable problems,'' in {\em Handbook of Mathematical Logic}
  (J.~Barwise, ed.), pp.~567--594, North-Holland, Amsterdam, 1977.

\end{thebibliography}
\bibliographystyle{ieeetr}

\newpage
\appendix

\section{ Proof of Lemma \ref{dobre-liczenie} }

We first deal with the equality for $\pi_s$. Suppose $D$ is a correct database.
In order to count the homomorphisms $h:\pi_s\rightarrow D$ let us group them by $h(x)$, that is by the vertex of $D$ the variable $x$
is mapped to.

Clearly, for $h$ to be a homomorphism $h(x)$ must be one of the constants $a_m$ for some $m\in \{1,\ldots \mmm\}$ (this is the only way the
atoms of the relations $R_d$ from $\pi_x$ can possibly be satisfied in $D$). So let us fix an $m$ such that $h(x)=a_m$.

The Lemma will follow once we are able to prove that:
~
$$ |\{h\in Hom(\pi_s, D): h(x)=a_m  \}| = c_{s,m}t_m(\Xi_D(\vec \xxi))   \;\;\; (*)   $$

Let us now split $\pi_s$ into $\pi'_s$ and $\pi''_s$, where:\bigskip

$$\pi'_s=   \bigwedge_{d \in \{1,\ldots \ddd\} }\; R_d(x,y_d)\wedge X(y_d,z_d)  $$\bigskip

$$\pi''_s=  \;\bigwedge_{m \in \{1,\ldots \mmm\} }\; S_{m}(x,x)\wedge S_{m}(x,x^m_{c_{s,m}})\;\;\wedge \bigwedge_{m \in \{1,\ldots \mmm\} }\;\bigwedge_{ 1\leq k < c_{s,m}}\; S_{m}(x^m_{k+1},x^m_{k})$$

Since $\pi'_s$ and  $\pi''_s$ do not share variables (except for $x$), in order to prove (*) it will be enough to show that:
~
$$ \hspace{30mm} |\{h\in Hom(\pi'_s, D): h(x)=a_m  \}| = t_m(\Xi_D(\vec \xxi))   \hspace{30mm}  \text{(**)}   $$

and that:
~
$$ \hspace{33mm}  |\{h\in Hom(\pi''_s, D): h(x)=a_m  \}| = c_{s,m}   \hspace{33mm}  \text{(***)}   $$

To see (**) notice that  for each $d\in \{1,\ldots \ddd\}   $   there is a unique way for $h(y_d)$ to be defined. But, again for each
$d\in \{1,\ldots \ddd\}   $, the value of  $h(z_d)$ can be chosen in $\Xi_D(\xxi_{m_d})$ ways, where $\xxi_{m_d}$ is the $d$-th variable of the term
$t_m$. And the choices, for each $d$, are independent, so indeed:
~
$$ |\{h\in Hom(\pi'_s, D): h(x)=a_m  \}| = \Pi_{d\in \{1,\ldots \ddd\}  }  \Xi_D(\xxi_{m_d}) =  t_m(\Xi_D(\vec \xxi))  $$

Now, let us concentrate on (***). Recall that $\pi''_s$ is a star, with $x$ as its center, and (for each $j\in \{1,\ldots \mmm\}$ )  with a
``ray'' consisting of $c_{s,j}-1$ edges of relation $S_j$ radiating from this center. On the other hand, in our correct database $D$,
there is a loop $S_j(a_m, a_m)$ for each $j\in \{1,\ldots \mmm\}$  (recall that $h(x)=a_m$) and there are no other edges
of the form  $S_j(a_m, \_)$ for $j\neq m$. This means that if $j\neq m$ then the $S_j$-ray can only be mapped to $D$ in one way: by mapping
all its edges to $S_j(a_m, a_m)$. The situation is different when $j=m$. Then we still have the loop $S_m(a_m, a_m)$ in $D$, but there is
also  $S_m(a_m, a)$ and  $S_m(a, a)$. This means that the $S_m$-ray can again  be mapped to $D$  by mapping
all its edges to $S_m(a_m, a_m)$. But it can also loop in $a_m$ for some time, then have one of its edges mapped to $S_m(a_m, a)$ and then
have all the remaining edges mapped to   $S_m(a, a)$. The edge to be mapped to $S_m(a_m, a)$ can be chosen in $c_{s,j}-1$ ways, and hence we get (***).

The argument for $\pi_b$ is almost analogous. The only difference is that now we need to notice that:
~
$$\Xi_D(\xxi_1)^\ddd = (\bigwedge_{d \in \{1,\ldots \ddd\} }\; R_1(x,y'_d)\wedge X(y'_d,z'_d)\;\;\;)(D)   $$
\qed

\section{ Proof of Lemma \ref{source-of-und}}\label{proof-of-h10}

\newcommand{\mycircfull}{ \ding{108} }
\newcommand{\mycircempty}{ \ding{109} }

In this appendix we show how Lemma \ref{source-of-und} follows from undecidability of Hilbert's 10th problem.

\subsection{Hilbert's 10th problem}

Before we start, let us clarify what exactly we mean by Hilbert's 10th problem. In order to do it, we refer to the canonical resource \cite{Davis77}. One of the results presented there, the one we are interested in, can be formulated as:

\begin{theorem}{(Hilbert's 10th problem undecidablity)}\label{H10-original}
\medskip

{\em The problem:}

Given is polynomial $Q$ of numerical variables $\xi_{Q, 1}, ..., \xi_{Q, \nnn_Q}$ with integer coefficients.
Does it hold that $Q(\Xi(\bar{\xi_Q})) \neq 0$ for every valuation $\Xi:\{\xi_{Q, 1}, \xi_{Q, 2},\ldots  \xi_{Q, \nnn_Q}\}\rightarrow \mathbb N$? \medskip

{\em is undecidable.}

\end{theorem}

We are going to show how the above Theorem implies Lemma \ref{source-of-und}. As a way to do it, for any $Q$ being an instance of 10th Hilbert's problem (understood like in Theorem \ref{H10-original}), we will construct a triple $[P_s, P_b, \cccc]$ -- the corresponding instance of the problem from Lemma \ref{source-of-und}. Keeping the notation from Lemma \ref{source-of-und} and from Theorem \ref{H10-original}, the constructed instance should be such that these are equivalent: \medskip

\mycircfull there exists valuation $\Xi: \{\xi_{Q, 1}, ..., \xi_{Q, \nnn_q}\}\rightarrow \mathbb N$ such that
$$ Q(\Xi(\bar{\xi_Q})) = 0 $$

\mycircempty there exists valuation $\Xi': \{\xi_1, ..., \xi_{\nnn}\} \rightarrow \mathbb N$ such that
$$ \cccc P_s(\Xi'(\bar{\xi})) > \Xi'(\xi_1)^d P_b(\Xi'(\bar{\xi})) $$

\subsection{From one polynomial to two}

Let $Q$ be any polynomial like in Theorem \ref{H10-original}. Rename the variables of $Q$ as $\xi_2, ..., \xi_n$. From this point, we will  denote by $\bar{\xi}$ the tuple $[\xi_2, ..., \xi_n]$. Note that we start indexing from $2$ on purpose (in particular, $n = \nnn_Q + 1$, if we refer to the notation from Lemma \ref{H10-original}). The reason for such choice will become clear soon. \\

We will now construct two polynomials $P_1$ and $P_2$ of the same variables as $Q$, which both have natural coefficients and such that for any fixed valuation $P_1$ is greater than $P_2$ if and only if $Q$ equals zero for the same valuation.

First consider polynomial $Q' = Q^2$ It can of course be written in the form:
$$Q' = \Sigma_{i = 1}^{m_{Q'}} c_i t_i,$$

where $m_{Q'} \in \mathbb N$, the values $c_i \in \mathbb Z$ are coefficients and $t_i$ are monomials (of any degree each). Let now $I_+$ and $I_-$ be the sets of indices of positive and negative coefficients, respectively:
$$I_+ = \{1 \leq i \leq m_{Q'}: c_i > 0\}$$
$$I_- = \{1 \leq i \leq m_{Q'}: c_i < 0\}$$

This allows us to introduce the polynomials:
$$Q'_+ = \Sigma_{i \in I_+} c_i t_i$$
and
$$Q'_- = \Sigma_{i \in I_-} (-c_i) t_i$$

It is easy to see that $Q' = Q'_+ - Q'_-$ and that all the coefficients in $Q'_+$ and in $Q'_-$ are natural. Let $P_1 = Q_- + 1$ and $P_2 = Q_+$. Observe that:

\begin{lemma}\label{from-q-to-p}
For each valuation $\Xi : \{\xi_2, ..., \xi_n\}\rightarrow \mathbb N$ it holds that
$Q(\Xi(\bar{\xi})) = 0$ if and only if $P_1(\Xi(\bar{\xi})) > P_2(\Xi(\bar{\xi}))$.
\end{lemma}

\noindent {\em Proof:~} For any valuation $\Xi$ like above we have: \medskip

$Q(\Xi(\bar{\xi})) = 0 \Leftrightarrow \\
\Leftrightarrow Q^2(\Xi(\bar{\xi})) = Q'(\Xi(\bar{\xi})) < 1 \Leftrightarrow \\
\Leftrightarrow Q'_+(\Xi(\bar{\xi})) - Q'_-(\Xi(\bar{\xi})) < 1 \Leftrightarrow \\
\Leftrightarrow Q'_+(\Xi(\bar{\xi})) < Q'_-(\Xi(\bar{\xi})) + 1 \Leftrightarrow \\
\Leftrightarrow P_2(\Xi(\bar{\xi})) < P_1(\Xi(\bar{\xi}))$. \qed \\

Lemma \ref{from-q-to-p} is the first span in our bridge between Theorem \ref{H10-original} and Lemma \ref{source-of-und}. But it is not the entire bridge yet,
recall that Lemma \ref{source-of-und} mentions several conditions which $P_s$ and $P_b$ have to satisfy. And, clearly, there is no reason to think that
$P_1$ and $P_2$ will always satisfy all of them.

 So now we will, step by step, keep adjusting $P_1$ and $P_2$ to consecutive conditions from Lemma \ref{source-of-und}. Finally, we will obtain a pair of polynomials which indeed can be taken as $P_s$ and $P_b$ (the constant $\cccc$ will appear in the meantime).

At the end, we want to refer back to Lemma \ref{from-q-to-p}. In order to be able to do this, we need to construct $P_s$ and $P_b$ in such a way that they in some sense inherit the initial inequality between $P_1$ and $P_2$.

\subsection{Common monomials}

We can express $P_1$ and $P_2$ in the form similar to $Q'$, that is:
$$P_1 = \Sigma_{i = 1}^{m_1} c_{1, i} t_{1, i}$$
$$P_2 = \Sigma_{i = 1}^{m_2} c_{2, i} t_{2, i}$$

where $m_1, m_2 \in \mathbb N$, $c_{j, i} \in \mathbb N$ are again coefficients and $t_{j, i}$ are again monomials.

Denote by $T_1$ the set of monomials of $P_1$, that is $\{t_{1, 1}, ..., t_{1, m_1}\}$. Analogously, let $T_2$ be such set for $P_2$. If we wanted to take $P_1$ and $P_2$ as $P_s$ and $P_b$, respectively, then one of the conditions from Lemma \ref{source-of-und} can be expressed as $T_1 = T_2$.

However, it may happen that $T_1 \neq T_2$. To deal with this problem, set $T = T_1 \cup T_2$, that is, the set of monomials appearing in $P_1$ or in $P_2$. For further convenience, it is helpful to denote the monomials belonging to $T$ as $t_1, ..., t_m$, where $m = |T|$. It is also useful to introduce the polynomial $P = \Sigma_{i = 1}^m t_i$. Now, instead of $P_1$ and $P_2$, consider:
$$P_1' = P_1 + P$$ and $$P_2' = P_2 + P$$

These polynomials have common set of monomials, namely $T$. This allows us to express them in the form we are already familiar with:
$$P_1' = \Sigma_{i = 1}^m c'_{1, i} t_i$$
$$P_2' = \Sigma_{i = 1}^m c'_{2, i} t_i$$

Note that the new coefficients $c'$ are different then the ones appearing in $P_1$ and $P_2$, but they still are natural numbers.

\subsection{Common degree and the  variable $\xi_1$}

The next problem is that we want all  $t_i \in T$ to be of the same degree degree. Moreover, we want all $t_i$ to start with the same variable. Let us try to deal with these both constraints simultaneously, introducing the new variable $\xi_1$.

Denote by $d_i$ the degree of monomial $t_i$. Take $d = 1 + max \{d_1, ...,  d_m\}$ and define:
$$t_i' = \xi_1^{d - d_i} t_i.$$

Now, define $P''_1, P''_2$ to be:

$$P_1'' = \Sigma_{i = 1}^m c'_{1, i} t_i'$$
$$P_2'' = \Sigma_{i = 1}^m c'_{2, i} t_i'$$

Notice that the coefficients remained unchanged in $P_1''$ and $P_2''$ (compared to $P_1'$ and $P_2'$), we just modified the monomials a little bit, so that
they are all now divisible by $\xi_1$ and all have  degree $d$.

Once we defined $P_1''$ and $P_2''$, one can notice that:

\begin{lemma}\label{p-bis-properties} For any valuation
$\Xi: \{ \xi_2, ..., \xi_n \} \rightarrow \mathbb N$ and $a \in \mathbb N$ it holds that:

\begin{itemize}
\item  $P_1''([1, \Xi(\bar{\xi})]) = P_1'(\Xi(\bar{\xi}))$ 
\item $P_1''([a, \Xi(\bar{\xi)}]) \leq a^d P_1'(\Xi(\bar{\xi})) \leq a^d P_1''(\Xi(\bar{\xi}))$
\item the two above claims hold also for $P_2''$ and $P_2'$, respectively 
\end{itemize}
\end{lemma}

\noindent {\em Proof:~} The first claim is a trivial consequence of the fact that to construct $P_1''$ from $P_1'$, we only multiply each of the terms appearing in $P_1'$ by some power of $\xi_1$.

To make sure the second claim holds, it is convenient to consider each of the inequalities separately. For the first inequality, it is enough to notice that for each $1 \leq i \leq m$ we have:
$$c'_{1, i} t_i'([a, \Xi(\bar{\xi})]) = c'_{1, i} a^{d - d_i} t_i(\Xi(\bar{\xi})) \leq a^d c'_{1, i} t_i(\Xi(\bar{\xi}))$$

The equality above follows from the definition of $t_i'$.

The prove the above inequality  two cases need to be considered. The simpler one is when $a = 0$, as the inequality is trivially satisfied then. When $a > 0$, for each $1 \leq i \leq m$ we have:
$$a^{d - d_i} \geq 1$$

which implies:
$$c'_{1, i} t_i(\Xi(\bar{\xi})) \leq c'_{1, i} t_i'([a, \Xi(\bar{\xi})])$$

The second second inequality becomes clear after multiplying both sides by $a^d$.

The last claim can be shown by repeating exactly the same arguments for $P_2''$ and $P_2'$. \qed \\

Now we want to show that the inequality between $P_1$ and $P_2$ in some sense survives when we consider $P_1''$ and $P_2''$ instead of them. The exact way it happens is presented by the next two lemmas.

\begin{lemma}\label{ksi-1-rowne-1}
Let $\Xi : \{\xi_2, ..., \xi_n\} \rightarrow \mathbb N$ be any valuation such that $P_1(\Xi(\bar{\xi})) > P_2(\Xi(\bar{\xi}))$. Then $P_1''([1, \Xi(\bar{\xi})]) > P_2''([1, \Xi(\bar{\xi})])$.

\end{lemma}

\noindent {\em Proof:~}

$P_1''([1, \Xi(\bar{\xi})]) = P_1'(\Xi(\bar{\xi})) = P_1(\Xi(\bar{\xi})) + P(\Xi(\bar{\xi})) > $ \\
$> P_2(\Xi(\bar{\xi})) + P(\Xi(\bar{\xi})) = P_2'(\Xi(\bar{\xi})) = P_2''([1, \Xi(\bar{\xi})])$\smallskip

The first and last equalities come from the first and third claims of Lemma \ref{p-bis-properties}. \qed 

\begin{lemma}\label{ksi-1-rowne-a}
Let $\Xi' : \{\xi_1, ..., \xi_n\} \rightarrow \mathbb N$ be any valuation. Consider the valuation $\Xi : \{\xi_2, ..., \xi_n\} \rightarrow \mathbb N$ such that $\Xi'([\xi_1, \bar{\xi}]) = [a, \Xi(\bar{\xi})]$ for some $a \in \mathbb N$. If additionally $P_1(\Xi(\bar{\xi})) \leq P_2(\Xi(\bar{\xi}))$ holds, then also $P_1''([a, \Xi(\bar{\xi})]) \leq a^d P_2''([a, \Xi(\bar{\xi})])$.

\end{lemma}

\noindent {\em Proof:~} $P_1''([a, \Xi(\bar{\xi})]) \leq a^d P_1'(\Xi(\bar{\xi})) = a^d (P_1(\Xi(\bar{\xi})) + P(\Xi(\bar{\xi}))) \leq \\
\leq a^d (P_2(\Xi(\bar{\xi})) + P(\Xi(\bar{\xi}))) =  a^d P_2'(\Xi(\bar{\xi})) \leq a^d P_2''([a, \Xi(\bar{\xi})])$.

Here the first and last inequalities come from the second and third claims of Lemma \ref{p-bis-properties}. \qed

\subsection{The final construction}

$P_1''$ and $P_2''$ are almost what we want to have as $P_s$ and $P_b$. Almost, because we are still left with the last condition from Lemma \ref{source-of-und}: for each $1 \leq i \leq m$ we require $c'_{1, i} \leq c'_{2, i}$ to hold.

This is the place where the constant $\cccc$ plays its role: to enforce inequalities mentioned in the previous sentence, we may multiply $P_2$ by some big number. It is enough to take:
$$\cccc = max \{c'_{1, 1}, c'_{1, 2}, ..., c'_{1, m}\}$$

We may thus finally define also:
$$P_s = P_1'' \text{~~~ and ~~~ } P_b = \cccc P_2''$$

The last step is to notice that indeed for each $1 \leq i \leq m$ we have now:
$$c_{s, i} = c'_{1, i} \leq \cccc \leq \cccc c'_{2, i} = c_{b, i}$$
where $c_{s, i}$ and $c_{b, i}$ are understood like in Lemma \ref{source-of-und}.

\subsection{Proof of the equivalence}

If we carefully followed the construction of $P_s$ and $P_b$, we should be now confident that they (alongside with $\cccc$) satisfy all the conditions from Lemma \ref{source-of-und}. Thus, we just need to prove the following lemma, which together with Lemma \ref{from-q-to-p} will tell us that \mycircfull $\Leftrightarrow$ \mycircempty:

\begin{lemma}

The following conditions are equivalent:

\noindent
$\bullet$~ there exists valuation $\Xi : \{\xi_2, ..., \xi_n\} \rightarrow \mathbb N$ such that  \hspace{1mm}
$P_1(\Xi(\bar{\xi})) > P_2(\Xi(\bar{\xi}))$;

\noindent
$\bullet$~  there exists valuation $\Xi' : \{\xi_1, \xi_2, ..., \xi_n\} \rightarrow \mathbb N$ such that \hspace{1mm}
$\cccc P_s(\Xi'([\xi_1, \bar{\xi}])) > \Xi'(\xi_1)^d P_b(\Xi'([\xi_1, \bar{\xi}]))$.

\end{lemma}

\noindent {\em Proof:~} Consider first the situation when for some valuation $\Xi$ we have $P_1(\Xi(\bar{\xi})) > P_2(\Xi(\bar{\xi}))$. Take $\Xi'$ such that:
$$\Xi'([\xi_1, \bar{\xi}]) = [1, \Xi(\bar{\xi})].$$
From Lemma \ref{ksi-1-rowne-1} we can conclude that:
$$P_1''([1, \Xi(\bar{\xi})]) > P_2''([1, \Xi(\bar{\xi})])$$ 
and multiplying both sides by $\cccc$ allows us to write: \medskip

$\cccc P_s(\Xi'([\xi_1, \bar{\xi}])) = \cccc P_s([1, \Xi(\bar{\xi})]) = \cccc P_1''([1, \Xi(\bar{\xi})]) > \\
> \cccc P_2''([1, \Xi(\bar{\xi})]) = P_b([1, \Xi(\bar{\xi})]) = P_b(\Xi'([\xi_1, \bar{\xi}]))
= \Xi'(\xi_1)^d P_b(\Xi'([\xi_1, \bar{\xi}]))$. \medskip

In the last equality, we were allowed to insert $\Xi'(\xi_1)^d$ before $P_b(\Xi'([\xi_1, \bar{\xi}]))$, as $\Xi'(\xi_1) = 1$. \\

Now assume that $P_1(\Xi(\bar{\xi})) \leq P_2(\Xi(\bar{\xi}))$ holds for all valuations $\Xi$. Consider then any valuation $\Xi'$, we know that there exist $a \in \mathbb N$ and valuation $\Xi$ such that \hspace{0.5mm}  $\Xi'([\xi_1, \bar{\xi}])= [a, \Xi(\bar{\xi})]$.
Lemma \ref{ksi-1-rowne-a}  tells us that we then have \hspace{0.5mm}  $P_1''([a, \Xi(\bar{\xi})]) \leq a^d P_2''([a, \Xi(\bar{\xi})])$.
Let us again multiply both sides by $\cccc$, obtaining in this case: \medskip

$\cccc P_s(\Xi'([\xi_1, \bar{\xi}])) = \cccc P_s([a, \Xi(\bar{\xi})]) = \cccc P_1''([a, \Xi(\bar{\xi})]) \leq \\
\leq \cccc a^d P_2''([a, \Xi(\bar{\xi})]) = a^d P_b([a, \Xi(\bar{\xi})]) = \Xi'(\xi_1)^d P_b(\Xi'([\xi_1,  \bar{\xi}]))$. \qed


\end{document}